\documentclass[prb,superscriptaddress,preprintnumbers,amsmath,amssymb]{revtex4}
\usepackage{epsfig}
\usepackage{verbatim}
\usepackage{color}
\usepackage{amsmath}

\usepackage{longtable}
\usepackage{hyperref}

\begin{document}
\title{Extensions of the time-dependent density functional based tight-binding approach}

 \author{A. Dom\'inguez}
 \affiliation{Bremen Center for Computational Materials Science, Universit\"at Bremen, Am Fallturm 1, 28359 Bremen, Germany}
 \author{B. Aradi}
 \affiliation{Bremen Center for Computational Materials Science, Universit\"at Bremen, Am Fallturm 1, 28359 Bremen, Germany}
 \author{T. Frauenheim}
 \affiliation{Bremen Center for Computational Materials Science,
   Universit\"at Bremen, Am Fallturm 1, 28359 Bremen, Germany}
 \author{V. Lutsker}
 \affiliation{Department of Theoretical Physics, University of Regensburg, 93040 Regensburg, Germany}
 \author{T. A. Niehaus}
 \affiliation{Department of Theoretical Physics, University of Regensburg, 93040 Regensburg, Germany}
\email{adrielg@bccms.uni-bremen.de}



\begin{abstract}
The time-dependent density functional based
tight-binding (TD-DFTB) approach is generalized to account for
fractional occupations. In addition, an on-site correction leads to
marked qualitative and quantitative improvements over the original method. Especially, the known failure
of TD-DFTB for the description of $\sigma \rightarrow \pi^*$ and $n
\rightarrow \pi^*$ excitations is overcome. Benchmark calculations on
a large set of organic molecules also indicate a better description of
triplet states. The accuracy of the revised TD-DFTB method is found to
be similar to first principles TD-DFT calculations at a highly reduced
computational cost.   As a side issue, we also discuss the generalization of the TD-DFTB method to spin-polarized systems. In contrast to an earlier study [Trani et al., JCTC {\bf 7} 3304 (2011)], we obtain a formalism that is fully consistent with the use of local exchange-correlation functionals in the ground state DFTB method.

\end{abstract}

\maketitle

\clearpage

\section{INTRODUCTION}

During the last years, Density Functional Theory (DFT) has been  one
of the most utilized tools for the description of ground-state
properties of a wide variety of molecular systems that range from
small molecules to large periodic materials. While it lacks the
accuracy typical of correlated wavefunction-based methods, it goes
beyond Hartree-Fock (HF) as electron correlation is incorporated in a
self-consistent-field (SCF) fashion. Thus, DFT has turned out to be a
good compromise between accuracy and computational cost; affordable to
study hundreds-of-atoms systems on most of the current work stations
with fairly good precision. The field of application of this method
was subsequently extended to the study of excited states properties
with the development of time dependent density functional
theory (TD-DFT).\cite{runge1984dft}  This method has become
the de facto standard for the computation of optical properties for
molecules with several tens of atoms. Also the limitations of TD-DFT
are now well documented in the literature (see
e.g. \cite{Casida2012}), which allows researchers to judge a priori
whether a certain class of exchange-correlation functionals is sufficient
for the predictive simulation of the problem at hand.

Still, many applications in photochemistry and nanophysics are not
easily tractable in the current methodological frame. As an example, quantum
molecular dynamics in the excited state require the evaluation of
energies and forces at a large number of points along the
trajectory. Also, the investigation of extended nanostructures like nanowires and -tubes
with intrinsic defects or surface modifications can not be reliably
performed with small models. For these kind of problems an approximate
TD-DFT method might be advantageous. Such a scheme is the time dependent density functional based tight
binding method (TD-DFTB).\cite{Niehaus2001a} In the TD-DFTB method
additional approximations beyond the choice of a given
exchange-correlation functional are performed to enhance the numerical
efficiency. These are mostly the neglect and simplification of
two-electron integrals occurring in the linear response formulation of
TD-DFT. In contrast to HF-based semi-empirical methods (like INDO/S or
CIS-PM3), TD-DFTB approximates a reference method that already covers electronic
correlation and does not include free parameters that are adjusted to
reproduce experimental data.

After the original linear response implementation,\cite{Niehaus2001a} TD-DFTB was
extended in a number of different directions.  Analytical excited
state gradients have been derived,\cite{heringer2007aes} as well as a real time propagation of
Kohn-Sham (KS) orbitals that enables the computation of optical
spectra using order-N algorithms.\cite{Yam2003}  Several groups worked on non-adiabatic molecular
dynamics simulations in the Ehrenfest \cite{Niehaus2005} or surface
hopping \cite{Mitric2009,Jakowski2012}
approach. More recent developments include a TD-DFTB approach for open
boundary conditions in the field of quantum transport,\cite{Wang2011} as well as an implementation
for open shell systems.\cite{Trani2011} A detailed review that summarizes advantages
and limits of the method is also
available.\cite{Niehaus2009}

One of these limits is the rather poor description of $\sigma
\rightarrow \pi^*$ and $n \rightarrow \pi^*$ excitations in many
chromophores. In TD-DFTB these transitions show vanishing oscillator
strengths although higher levels of theory clearly predict a finite
(albeit small) transition probability. Moreover, singlet-triplet gaps
for these kind of excitations are exactly zero in TD-DFTB, which is
again in disagreement with more accurate methods. Although $\pi
\rightarrow \pi^*$ transitions usually dominate the absorption
spectrum, such low-lying excitations may play a significant
role in the luminescence properties that are of key importance in
many technological applications.

In this paper, we present extensions of the TD-DFTB method in order to improve its description of the absorption spectra of molecules. In the first section, the spin-unrestricted DFTB formulation for the ground state is briefly reviewed. Afterwards, we generalize the TD-DFTB approach for the study of open-shell systems with fractional occupancy. In \ref{onsite}, a refinement to overcome important deficiencies within the method is formulated. In order to highlight the qualitative improvement due to these extensions, we report results for selected diatomic molecules in \ref{results}. Additionally, a comparison between results obtained with the proposed formalism and the original TD-DFTB approach for a large set of benchmark molecules is presented. Our findings are further compared to TD-DFT, the best theoretical estimates from the literature and experimental observations.

\section{Spin-unrestricted DFTB for the ground state}
\label{spin-ground}
This section contains a brief summary of the spin-unrestricted DFTB
method in order to establish the required notation for the later
parts. More details on the derivation and application of the method are given in the
original article by K\"{o}hler and co-workers.\cite{Kohler2001}

DFTB is based on a expansion of the DFT-KS energy up to the second
order in the charge density fluctuations, $\delta \rho= \delta
\rho_{\uparrow} + \delta \rho_{\downarrow}$, around a reference
density, $\rho_0 = \left[\rho_{0 \uparrow}, \rho_{0 \downarrow
    }\right]$. The latter is given by a superposition of
pre-calculated densities
for neutral spin-unpolarized atoms. The DFTB energy reads:

\begin{eqnarray}
 E &=& \sum_{\sigma} \sum_i n_{i\sigma} \langle \psi_{i\sigma}
 |\hat{H}^0| \psi_{i\sigma} \rangle \nonumber\\
 &&+  \frac{1}{2} \sum_{\sigma \tau} \iint
  \delta \rho_{\sigma} (\mathbf{r})\, f^\text{hxc}_{\rho_\sigma \rho_{\tau}}[\rho_0](\mathbf{r}, \mathbf{r^{\prime}}) \, \delta \rho_{\tau} (\mathbf{r}^{\prime} ) \, d\mathbf{r} d\mathbf{r}^{\prime} +  E_\text{rep},
\label{Etot1}
\end{eqnarray}

where the $n_{i\sigma}$ denote occupation numbers, $E_\text{rep}$ is a
sum of pair potentials independent of the electronic
density,\cite{Frauenheim2002} and $f^\text{hxc} = f^\text{h} + f^\text{xc}$
with
\begin{eqnarray}
   f^\text{h}(\mathbf{r}, \mathbf{r^{\prime}})  &=&
   \frac{1}{|\mathbf{r} -
     \mathbf{r}^{\prime}|} \nonumber\\
 f^\text{xc}_{\rho_\sigma \rho_{\tau}}[\rho](\mathbf{r},
 \mathbf{r^{\prime}}) &=&  \frac{\delta^2 E_{xc}[\rho]}{\delta \rho_{\sigma} (\mathbf{r}) \delta \rho_{\tau}(\mathbf{r}^{\prime})},
\end{eqnarray}
denote the Coulomb and exchange-correlation kernel, respectively.

In the first (zeroth-order) term of \ref{Etot1}, $\hat{H}^0$ is
the KS Hamiltonian evaluated at $\rho_0$ and the sum runs over the
spin variables $\sigma = \uparrow, \downarrow$ and the KS orbitals
$\psi_{i\sigma}$. For the succeeding formulation, the Roman indices
$s,t,u,v,$ denote KS orbitals, whereas Greek indices
$\mu,\nu,\kappa,\lambda$ will denote atomic orbitals (AO). Let us also
abbreviate a general two-point integral over a kernel $g(\mathbf{r},
\mathbf{r^{\prime}})$ in the following form:

\begin{equation}
  \left( f | g | h \right) =   \iint f( \mathbf{r}) g(\mathbf{r},
\mathbf{r^{\prime}}) h(\mathbf{r^{\prime}}) \, d\mathbf{r} d\mathbf{r}^{\prime}.
\end{equation}

For atomic orbital products $f(\mathbf{r}) = \phi_\mu(\mathbf{r})
\phi_\nu(\mathbf{r})$ and $h(\mathbf{r}^{\prime}) = \phi_\kappa(\mathbf{r}^{\prime})
\phi_\lambda(\mathbf{r}^{\prime})$ the shorthand $ \left( \mu\nu
  | g | \kappa\lambda \right)$ will be also used. Expanding the KS orbitals
(which we choose to be real) into a suitable set of such localized
atom-centered AO, $\psi_{i\sigma} = \sum_{\mu} c_{\mu i}^{\sigma}
\phi_{\mu}$, the second term in \ref{Etot1}, labeled $E^{(2)}$ in
the following, can be written as

\begin{equation}
E^{(2)} = \frac{1}{2} \sum_{\sigma \tau} \sum_{\mu\nu\kappa\lambda} \Delta P_{\mu\nu}^{\sigma}  \left(\mu \nu | f^\text{hxc}_{\rho_\sigma \rho_\tau}[\rho_0]| \kappa \lambda \right)\Delta P_{\kappa\lambda}^{\tau},
\label{E2nd}
\end{equation}
where $\Delta P_{\mu\nu}^{\sigma} = P_{\mu\nu}^{\sigma} - P_{\mu\nu}^{0,\sigma}$ denotes the AO density matrix of the difference density $\delta \rho_\sigma(\mathbf{r})$, with

\begin{equation}
 P_{\mu\nu}^{\sigma} = \sum_{st} c_{\mu s}^{\sigma} P_{st}^{\sigma}
 c_{\nu t}^{\sigma}, ~~~ P_{\mu\nu}^{0,\sigma} = \sum_{st} c_{\mu
   s}^{\sigma} P_{st}^{0,\sigma}  c_{\nu t}^{\sigma}.
\label{AO_P}
\end{equation}
Here $ P_{st}^{\sigma} = \langle \Psi |\hat{a}_{t\sigma}^{\dagger} \hat{a}_{s\sigma}| \Psi \rangle$ are the MO density matrix elements, $\Psi$ being the ground state KS determinant. The term $P_{st}^{0,\sigma}$
designates the MO density matrix of the reference system.

To evaluate the appearing integrals, the Mulliken
approximation is applied. This amounts to set $\phi_{\mu} \phi_{\nu}
\approx \frac{1}{2} S_{\mu\nu} (|\phi_{\mu}|^2+|\phi_{\nu}|^2)$, using
the known AO overlap integrals $S_{\mu\nu}$.  Thus, the general four-center integrals are written in terms of two-center integrals,

\begin{equation}
  \label{mull}
  \left(\mu  \nu |
  f^\text{hxc}_{\rho_\sigma \rho_\tau}[\rho_0]|  \kappa  \lambda
\right)\approx \frac{1}{4} S_{\mu\nu} S_{\kappa\lambda} \sum_{\alpha
  \in \{ \mu,\nu\}}  \sum_{\beta
  \in \{\kappa,\lambda\}}
       \left(\alpha  \alpha |
  f^\text{hxc}_{\rho_\sigma \rho_\tau}[\rho_0]|  \beta  \beta
\right).
\end{equation}
By substituting \ref{mull} into \ref{E2nd}, the second order contribution to the energy reads,

\begin{equation}
E^{(2)} = \frac{1}{2} \sum_{\sigma \tau} \sum_{\mu\nu} \Delta\tilde{P}_{\mu\mu}^{\sigma}  \left(\mu \mu | f^\text{hxc}_{\rho_\sigma \rho_\tau}[\rho_0]| \nu \nu \right)\Delta \tilde{P}_{\nu\nu}^{\tau},
\label{E2nd_2}
\end{equation}
where $\tilde{P}_{\mu\nu}^{\sigma}$ are the elements of the dual density matrix defined as

\begin{equation}
 \tilde{\mathbf{P}} = \frac{1}{2} \left( \mathbf{P}\cdot\mathbf{S} + \mathbf{S}\cdot\mathbf{P} \right).
\label{dualP}
\end{equation}
By using \ref{AO_P}, the dual density matrix can be expressed with respect to the MO density matrix as
\begin{equation}
 \tilde{P}_{\mu\nu}^{\sigma} = \sum_{st} P_{\mu\nu}^{st\sigma} P_{st}^{\sigma}.
\label{dualP_2}
\end{equation}
The matrix $ P_{\mu\nu}^{st\sigma}$ represents the Kohn-Sham transition density for an excitation from orbital s to t, and is defined as follows:

\begin{equation}
 P_{\mu\nu}^{st\sigma} = \frac{1}{4} \left( c_{\mu s}^{\sigma} \tilde{c}_{\nu t}^{\sigma} +  c_{\mu t}^{\sigma} \tilde{c}_{\nu s}^{\sigma} + c_{\nu s}^{\sigma} \tilde{c}_{\mu t}^{\sigma} + c_{\nu t}^{\sigma} \tilde{c}_{\mu s}^{\sigma} \right), ~~ \tilde{\mathbf{c}}_s =  \mathbf{c}_s\cdot\mathbf{S}.
\label{TDM}
\end{equation}

A further simplification to \ref{E2nd_2} is obtained by spherical averaging over AO
products. This ensures that the final total energy expression is invariant
with respect
to arbitrary rotations of the molecular frame. To this end, the functions $F_{Al}$ are introduced
  \begin{equation}
    \label{sph}
    F_{Al}( |{\bf r} - \mathbf{R}_A|) = \frac{1}{2l+1} \sum_{m=-l}^{m=l}|\phi_\mu({\bf r} - \mathbf{R}_A)|^2,
  \end{equation}
where $l$ and $m$ denote the angular momentum and magnetic quantum
number of AO $\phi_\mu$, centered on atom $A$. We then have for $\mu=\{Alm\}$, $\nu=\{Bl'm'\}$:
\begin{equation}
  \label{2cent}
  \left(\mu  \mu |
  f^\text{hxc}_{\rho_\sigma \rho_\tau}[\rho_0]|  \nu  \nu
\right)\approx \left( F_{Al}|
  f^\text{hxc}_{\rho_\sigma \rho_\tau}[\rho_0]|F_{Bl'}\right)  =  \Gamma_{Al, Bl^{\prime}}^{\sigma\tau}
\end{equation}
introducing the shorthand notation $\Gamma$.

The traditional DFTB energy expression is now obtained after transformation
from the set $\{\rho_\uparrow, \rho_\downarrow\}$ to the total density
$\rho= \rho_\uparrow+ \rho_\downarrow$ and magnetization $m=
\rho_\uparrow- \rho_\downarrow$. By this change of variables, the
exchange-correlation kernel can be split and one arrives at

\begin{equation}
 \Gamma_{Al, Bl'}^{\sigma\tau}  = \gamma_{Al,Bl'} + \delta_{\sigma}\delta_{\tau} \delta_{AB} W_{Al,l'},
\end{equation}
where $\delta_{\sigma} = 2\delta_{\sigma\uparrow} - 1$ and the parameters $\gamma_{Al,Bl'} = \left( F_{Al}|f^\text{hxc}_{\rho\rho}[\rho_0]| F_{Bl'} \right)$ and $W_{Al,l'} = \left( F_{Al}|f^\text{xc}_{mm}[\rho_0]| F_{Al'} \right)$. Please note that the constants $W_{Al,l'}$ depend
only on the exchange-correlation kernel, but not on the long-range
Coulomb interaction. Moreover, as the reference density $\rho_0$ is built from neutral
spin-unpolarized atomic densities, there are no integrals which
involve mixed derivatives of the exchange-correlation energy with
respect to both
density and magnetization.  The
parameters $\gamma_{Al,Bl'}$ and
$W_{Al,l'}$ are known in the DFTB
method as the $\gamma$-functional and spin coupling constants,
respectively.\cite{Frauenheim2002}  The spin coupling constants are treated as strictly
on-site parameters,
whereas $\gamma$ is calculated for every atom pair using an
interpolation formula, that depends on the distance between the atoms $A$
and $B$ and the atomic Hubbard-like parameters $\gamma_{Al,Al}$ and
$\gamma_{Bl,Bl}$. Traditionally, the latter are not computed directly
from the integral \ref{2cent}, but from total energy
derivatives according to 
$\gamma_{Al,Al} = \delta^2 E /\delta n^2$.
In this expression, $E$ denotes the DFT total energy of atom $A$
and $n$ refers to the occupation of the shell with angular momentum
$l$. The derivative is then evaluated numerically by full
self-consistent field calculations at perturbed
occupations. Due to orbital relaxation, parameters obtained in this
way are roughly 10-20 \% smaller than the ones from a direct integral
evaluation. This point will become important later.

Finally, by substituting \ref{dualP_2} in \ref{E2nd_2} while using \ref{2cent}, the second order energy term can be expressed as

\begin{equation}
E^{(2)} = \frac{1}{2} \sum_{\sigma \tau} \sum_{stuv} \sum_{ABll^{\prime}} \Delta P_{st}^{\sigma}
q_{Al}^{st\sigma}  \Gamma_{Al,Bl^{\prime}}^{\sigma \tau}q_{Bl^{\prime}}^{uv\tau}\Delta P_{uv}^{\tau} ,
\label{2nd_final}
\end{equation}
where $q_{Al}^{st\sigma} = \sum_{\mu \in A,l} P_{\mu\mu}^{st\sigma}$ are the angular-momentum-resolved transition charges for atom
$A$. For later reference, we next define a matrix $\bar{\mathbf{K}}$:

\begin{equation}
 \bar{K}_{st\sigma,uv\tau} = \sum_{ABll^{\prime}} q_{Al}^{st\sigma} \Gamma_{Al, Bl^{\prime}}^{\sigma\tau}q_{Bl^{\prime}}^{uv\tau} .
\label{coupling}
\end{equation}
Using this abbreviation, the DFTB total energy can then be written in the following simple form,

\begin{eqnarray}
 E &=& \sum_{\sigma} \sum_{st} H_{st\sigma}^0 P_{st}^{\sigma} +
 \frac{1}{2} \sum_{\sigma \tau} \sum_{stuv} \Delta P_{st}^{\sigma}
 \bar{K}_{st\sigma,uv\tau} \Delta P_{uv}^{\tau} +  E_\text{rep}.
\label{totE}
\end{eqnarray}
By applying the variational principle to the energy functional, the DFTB KS equations are obtained

\begin{equation}
H_{st\sigma} - \epsilon_{s\sigma} \delta_{st} = 0, ~~ \forall s,t,\sigma
\label{ks}
\end{equation}
where the KS Hamiltonian is expressed as

\begin{equation}
H_{st\sigma} = \frac{\partial E}{\partial P_{st}^{\sigma}} =
H_{st\sigma}^{0} + \sum_{\tau} \sum_{uv} \bar{K}_{st\sigma,uv\tau} \Delta P_{uv}^{\tau}.
\label{finalHam}
\end{equation}

These equations are subsequently transformed into a set of algebraic
equations by expanding the KS orbitals into the AO basis, and are
solved self-consistently. For the converged ground state density we
have $P_{st}^{\sigma} = n_{s\sigma} \delta_{st}$. Using also the identity
$\sum_{st} c_{\mu s}^\sigma P_{st}^{0,\sigma}  c_{\nu t}^\sigma = n^0_{\mu\sigma} \delta_{\mu\nu}~(\forall \mu,\nu,\sigma)$, $n^0_{\mu\sigma}$ being the occupation numbers for the reference atoms, it is straightforward to recover the expression for
the spin-unrestricted DFTB Hamiltonian given in \cite{Kohler2001,Frauenheim2002}.

\section{Spin-unrestricted TD-DFTB}
\label{spin}

Once the KS orbitals $\psi_{i\sigma}$ and energies
$\epsilon_{i\sigma}$ are obtained, a linear density-response treatment
directly applies as a natural extension to DFTB. Following Casida,\cite{Casida1995} the excitation energies $\omega_{I}$ and eigenvectors $F_{I}$ can be calculated by solving the eigenvalue problem

\begin{equation}
\boldsymbol{\Omega} \mathbf{F}_I = \omega_{I}^2 \mathbf{F}_I,
\label{omega}
\end{equation}
where the response matrix $\boldsymbol{\Omega}$ is defined as

\begin{equation}
 \boldsymbol{\Omega} = \mathbf{S}^{-1/2} \left( \mathbf{A} + \mathbf{B} \right) \mathbf{S}^{-1/2}
\end{equation}
\begin{equation}
 \mathbf{S} = -\mathbf{C} \left( \mathbf{A} - \mathbf{B} \right)^{-1} \mathbf{C}.
\end{equation}
The matrices $\mathbf{A}$, $\mathbf{B}$ and $\mathbf{C}$ are defined according to

\begin{eqnarray}
 A_{ia\sigma, jb\tau} &=& \frac{\delta_{ij} \delta_{ab} \delta_{\sigma \tau}\omega_{jb\tau}}{ n_{j\tau}-n_{b\tau}} + K_{ia\sigma,jb\tau}\nonumber\\
 B_{ia\sigma, jb\tau} &=&  K_{ia\sigma,bj\tau}\nonumber\\
 C_{ia\sigma, jb\tau} &=& \frac{\delta_{ij} \delta_{ab} \delta_{\sigma
     \tau}}{ n_{j\tau}-n_{b\tau}},
\label{ABC}
\end{eqnarray}
where $\omega_{jb\tau} = \epsilon_{b\tau} - \epsilon_{j\tau}$ and the
indexes $i,j,a,b$ stand for KS orbitals such that $n_{i\sigma} >
n_{a\sigma}$ and $n_{j\tau} > n_{b\tau}$. We explicitly allow for
fractional occupations at this point in order to be able to simulate
also metallic or near-metallic systems at elevated electronic
temperature.

The coupling matrix elements
$K_{ia\sigma, jb\tau}$ are generally defined as the derivative of the
KS Hamiltonian with respect to the density matrix elements. For the
special case of DFTB (see \ref{finalHam}), this leads to the matrix $\bar{\mathbf{K}}$ defined in \ref{coupling}

\begin{equation}
 K_{ia\sigma,jb\tau} := \frac{\partial H_{ia\sigma}} {\partial P_{jb}^{\tau}} = \bar{K}_{ia\sigma,jb\tau}.
\label{2nd_exp1}
\end{equation}
This quantity depends only on the $\gamma$ and $W$ parameters, as well
as on the Mulliken transitions charges obtained from the previous
ground state DFTB calculation.

An important feature of the coupling matrix for local or semi-local
exchange-correlation functionals is its invariance with respect to the
permutation of any connected (real orbital) indices
(e.g. $K_{ia\sigma,jb\tau} = K_{ia\sigma,bj\tau} =
K_{ai\sigma,jb\tau}$). This symmetry does not hold for functionals
involving Hartree-Fock exchange.\cite{Casida1995} In contrast to an earlier derivation,\cite{Trani2011} we find that the DFTB coupling matrix is in fact
symmetric. This
can be traced back to the transition density matrix, \ref{TDM}, that
evidently has this property. Since the ground state DFTB method is
derived as an approximation to local or semi-local DFT
only\footnote{See \cite{Niehaus2012} for an extension of DFTB to general
  hybrid functionals.}, this is an expected and necessary property and
implies that the orbital rotation Hessian $\mathbf{A}-\mathbf{B}$
becomes strictly diagonal. 
Thus, the expression for the response matrix elements is conveniently simplified to read

\begin{equation}
\Omega_{ia\sigma,jb\tau} =  \delta_{ij} \delta_{ab} \delta_{\sigma \tau} \omega_{jb\tau}^2 + 2 \sqrt{\left( n_{i\sigma}-n_{a\sigma} \right) \omega_{ia\sigma}} K_{ia\sigma, jb\tau} \sqrt{\left( n_{j\tau}-n_{b\tau} \right) \omega_{jb\tau}}.
\end{equation}

It is worth formulating the method for closed shell systems as
a particular case, in order to make contact with the original
derivation of TD-DFTB . In this special case the Mulliken
transition charges have the property
$q_{Al}^{ia\uparrow}=q_{Al}^{ia\downarrow}=q_{Al}^{ia}$. If, in
addition, the dependence of the $\gamma$-functional and $W$ constants
on the angular momentum is neglected, the coupling matrix simplifies to

\begin{equation}
 K_{ia\sigma,jb\tau} = \sum_{AB} q_{A}^{ia}  \left(  \gamma_{AB}
   +\delta_{\sigma}\delta_{\tau} \delta_{AB} W_{A} \right) q_{B}^{jb},
\end{equation}
with $q_{A}^{ia} = \sum_{l} q_{Al}^{ia}$, in
full agreement with the expression derived previously for spin-unpolarized
densities.\cite{Niehaus2001a}





Once the eigenvalue problem is solved within TD-DFTB, the oscillator strength related to excitation $I$ can be calculated as

\begin{equation}
f^I = \frac{2}{3} \sum_{k=1}^3 \left| \sum_{ia\sigma} \langle \psi_{i\sigma} | \hat{r}_k | \psi_{a\sigma} \rangle \sqrt{\left( n_{i\sigma}-n_{a\sigma} \right) \omega_{ia\sigma}} F_{ia\sigma}^I \right|^2,
\end{equation}
where $\hat{r}_k$ denotes the $k$-th component of the position
operator. The transition-dipole matrix elements are also subjected to
a Mulliken approximation ($\mathbf{R}_{A}$ is the position of atom $A$):

\begin{equation}
\label{dipmat}
 \langle \psi_{i\sigma} | \hat{ \mathbf{r}} | \psi_{a\sigma} \rangle
 \approx \sum_{A}  \mathbf{R}_{A} q_{A}^{ia\sigma}.
\end{equation}

\section{Beyond the Mulliken approximation}
\label{onsite}

While in general the Mulliken approximation accounts, at least
approximately, for the differential overlap of atomic orbitals, there
is an important exception. If orbitals $\phi_\mu$ and $\phi_\nu$ with
$\mu \neq \nu$
reside on the same atom, their product vanishes since the overlap
matrix reduces to the identity. As a consequence, the transition
charges are underestimated or often vanish identically for
excitations that feature localized MO, e.g. $n \to \pi^*$
transitions. This in turn means that the coupling matrix vanishes and
hence no correction of the ground state KS energy differences occurs
in the linear response treatment. More importantly, the
singlet-triplet gap is zero for such excitations.

A next level of approximation demands therefore the evaluation of all
one-center integrals of the exchange type, i.e., $\left(\mu  \nu | f^\text{hxc}_{\rho_\sigma \rho_\tau}[\rho_0]|  \mu  \nu \right)$ with $\mu \neq \nu$. This is how
Pople et al. proceeded in the development of the intermediate neglect
of differential overlap model (INDO) \cite{Pople1967} to overcome the
deficiencies encountered within the complete neglect of differential
overlap method (CNDO).\cite{Pople1965} Under the INDO model, the
one-center, two-electron integrals are fitted to atomic spectroscopic
data. In this work, the corresponding one-center integrals are calculated by numerical integration.

After inclusion of every onsite exchange-like integral within the DFTB formalism (see \ref{App_ener}), the second-order energy takes the new form:

 \begin{eqnarray}
 E^{(2)} &=& \frac{1}{2} \sum_{\sigma \tau} \sum_{A} \sum_{\mu\nu \in A} \Delta\tilde{P}_{\mu\mu}^{\sigma}  \left(\mu \mu | f^\text{hxc}_{\rho_\sigma \rho_\tau}[\rho_0]| \nu \nu \right) \Delta \tilde{P}_{\nu\nu}^{\tau}\nonumber \\
         &+& \sum_{\sigma\tau} \sum_{A} \sum_{\mu\nu \in A}^{\mu \neq
           \nu} \Delta\tilde{P}_{\mu\nu}^{\sigma}  \left(\mu \nu |
           f^\text{hxc}_{\rho_\sigma \rho_\tau}[\rho_0]| \mu \nu
         \right) \Delta\tilde{P}_{\mu\nu}^{\tau}\nonumber \\
         &+& \frac{1}{2} \sum_{\sigma \tau} \sum_{AB}^{A \neq B}
         \sum_{\mu \in A} \sum_{\nu \in B}
         \Delta\tilde{P}_{\mu\mu}^{\sigma}  \left(\mu \mu |
           f^\text{hxc}_{\rho_\sigma \rho_\tau}[\rho_0]| \nu \nu
         \right) \Delta \tilde{P}_{\nu\nu}^{\tau},
\label{E2nd_corri}
\end{eqnarray}
where the onsite integrals appear in the second term and the previous
energy expression (\ref{E2nd_2}) has been split into on-site (first
term) and off-site (third term) contributions.

By substituting \ref{dualP_2} in \ref{E2nd_corri} while using \ref{2cent} for the off-site component, $E^{(2)}$ can be finally written as

\begin{equation}
 E^{(2)} = \frac{1}{2} \sum_{\sigma\tau} \sum_{stuv} \Delta
 P_{st}^{\sigma}  K_{st\sigma,uv\tau} \Delta P_{uv}^{\tau},
\label{E2nd-Onsite}
\end{equation}
with the refined coupling matrix $\mathbf{K}$ being expressed as
\begin{eqnarray}
 K_{st\sigma,uv\tau} &=& \sum_{A} \sum_{\mu\nu \in A} P_{\mu\mu}^{st\sigma} \left(\mu \mu | f^\text{hxc}_{\rho_\sigma \rho_\tau}[\rho_0]| \nu \nu \right) P_{\nu\nu}^{uv\tau} \nonumber \\
                     &+& 2 \sum_{A} \sum_{\mu\nu \in A}^{\mu \neq \nu} P_{\mu\nu}^{st\sigma}  \left(\mu \nu | f^\text{hxc}_{\rho_\sigma \rho_\tau}[\rho_0]| \mu \nu \right) P_{\mu\nu}^{uv\tau}\nonumber \\
                     &+& \sum_{ABll'}^{A \neq B} q_{Al}^{st\sigma}
                     \Gamma_{Al,Bl'}^{\sigma\tau} q_{Bl'}^{uv\tau} .
\label{Kmatrix}
\end{eqnarray}

With inclusion of the exchange integrals, the spherical averaging over
AO products described in \ref{2cent} will in general not lead to an
expression which is invariant under a rotation of the atomic
axes. This has been discussed by Figeys et. al,\cite{Figeys1977} who
analyzed the rotational invariance (RI) of the INDO model. Taking {\em
  p}-type orbitals as an example, the following identity must hold to
preserve RI:
\begin{equation}
 (pp | f^\text{hxc}_{\rho_\sigma \rho_\tau}[\rho_0]| pp) - (pp | f^\text{hxc}_{\rho_\sigma \rho_\tau}[\rho_0]| p'p') = 2 (pp' | f^\text{hxc}_{\rho_\sigma \rho_\tau}[\rho_0]| pp'),~ \forall p,p'\,.
\label{RI}
\end{equation}
In the original DFTB formulation this requirement is fulfilled, as both integrals on the left-hand side
of \ref{RI} are approximated to have the same value, $\Gamma_{Ap,Ap}^{\sigma\tau}$,
while the integral on the right-hand side is neglected. To retain RI
within the present scheme,  one could  evaluate all on-site integrals
exactly so that \ref{RI} holds automatically. As detailed in
\ref{spin-ground}, the corresponding averaged $\Gamma$-parameters
would then differ from the ones usually employed in
ground state DFTB simulations, which are obtained from total energy
derivatives. Instead, we use  \ref{RI} to calculate all
required parameters, taking the exact exchange integrals and the
traditional $\Gamma$-parameters as input.\footnote{ For  {\em
  p}-type orbitals we have for example $ (pp |
f^\text{hxc}_{\rho_\sigma \rho_\tau}[\rho_0]| pp)  =
\Gamma_{Ap,Ap}^{\sigma\tau} + \frac{4}{3}(pp' |
f^\text{hxc}_{\rho_\sigma \rho_\tau}[\rho_0]| pp')$ and  $(pp |
f^\text{hxc}_{\rho_\sigma \rho_\tau}[\rho_0]| p'p') = 
\Gamma_{Ap,Ap}^{\sigma\tau}  - \frac{2}{3}(pp' |
f^\text{hxc}_{\rho_\sigma \rho_\tau}[\rho_0]| pp')$. Here the
$\Gamma$-parameters are obtained from total energy derivatives, while
the exchange integrals are evaluated directly from the wavefunction.}
In this way the modifications of the original method are kept as small
as possible, while RI is still exactly preserved.

The eqs \ref{E2nd-Onsite} and \ref{Kmatrix} represent a correction to the
conventional DFTB energy expression, whose implication for various
ground state properties certainly warrants a deeper investigation. Since
we are mainly interested in an improvement of the TD-DFTB scheme at
this point, we keep with the traditional DFTB scheme for the ground
state and employ the modified coupling matrix only in the response
part of the calculation, i.e. in \ref{ABC}.      

In a similar manner as the refinement of the coupling matrix was
achieved, the approximation for the dipole matrix elements
eq \ref{dipmat}, and hence the oscillator strength, can be improved by including all non-vanishing one-center dipole integrals (see \ref{App_dip}),

\begin{equation}
 \langle \psi_{i\sigma} | \mathbf{\hat{r}} | \psi_{a\sigma} \rangle = \sum_{A} \mathbf{R}_{A} q_{A}^{ia\sigma} + \sum_A \sum_{\mu\nu \in A}^{\mu \neq \nu} P_{\mu\nu}^{ia\sigma}  \langle \mu | \mathbf{\hat{r}} | \nu\rangle.
\end{equation}

According to the dipole selection rules only $s$-$p$ and $p$-$d$ dipole integrals are non-zero. Among the $s$-$p$ integrals, only those of the type $D_{sp}=  \langle s | r_k | p_k\rangle$ do not vanish, all of them being equal. With regards to the $p$-$d$ integrals, eleven of them are non-vanishing:

\begin{eqnarray}
 D_{pd} &=&  \langle p_y | x | d_{xy} \rangle = \langle p_x | y | d_{xy} \rangle = \langle p_y | z | d_{yz} \rangle = \langle p_z | y | d_{yz} \rangle = \langle p_z | x | d_{xz} \rangle = \langle p_x | z | d_{xz} \rangle \nonumber \\
  D_{pd} &=& \langle p_x | x | d_{x^2-y^2} \rangle = -\langle p_y | y | d_{x^2-y^2} \rangle \nonumber \\
 D_{pd}^{\prime} &=&\langle p_y | y | d_{z^2} \rangle = \langle p_x | x | d_{z^2} \rangle  \nonumber \\
  D_{pd}^{\prime\prime} &=& \langle p_z | z | d_{z^2} \rangle .
\end{eqnarray}

We would like to stress that the required additional atomic integrals in our
correction to the coupling matrix and oscillator strength are neither
freely adjustable parameters nor fitted to experiment. Instead, they are
calculated numerically with a special-purpose DFT code that was
recently developed in our group.   

Like in earlier studies with the DFTB method, the
Perdew-Burke-Ernzerhof (PBE) \cite{perdew1996gga} functional was used
in the calculation of Hamiltonian and overlap matrices, as well as the
computation of reference densities and basis functions.\footnote{The
  employed Slater-Koster tables (mio-0-1 \cite{elstner1998scc,Niehaus2001}) are available at \url{http://www.dftb.org/parameters/download/mio/mio_0_1}.} This also
holds for the new parameters, which are calculated once for every atom
type, stored in a file, and read when needed during the
calculation. 

The presented extensions of the TD-DFTB method have been implemented in a
development version of the DFTB+ code.\cite{Aradi2007a}

\section{RESULTS}
\label{results}
\subsection{Diatomic molecules}
To show the performance of our corrections to the coupling and dipole
matrices (hereafter referred to as the onsite correction), we
calculated the low-lying vertical excitation energies and their
corresponding oscillator strengths of three diatomic molecules for
which the improvements over traditional TD-DFTB are especially
noticeable. It is important to point out that only valence excited
states can be treated within TD-DFTB due to the employed minimal basis
set. \ref{diatomic} shows the excitation energies of NO, N$_2$
and O$_2$ calculated within both the corrected and the original
formulation of TD-DFTB. For NO and O$_2$, spin-unrestricted TD-DFTB
calculations have been performed, where doublet and triplet ground
states have been considered, respectively. In order to identify the
excited state multiplicity of the open-shell systems, the
expectation value of the square of the total spin operator, $\langle
S^2 \rangle$, is evaluated. We apply the expression derived in
Ref. \cite{Maurice1995} for unrestricted configuration interaction
singles (UCIS) wave functions for this purpose. In
\ref{diatomic}, a multiplicity is only assigned to those excited
states with low spin contamination. This covers the most important excitations in the absorption spectrum. 

As a reference, we computed
vertical excitation energies to valence states of these compounds by
using TD-DFT as implemented in the TURBOMOLE package
\cite{TURBOMOLE}. The PBE exchange-correlation functional as well as
triple-zeta plus polarization (TZP) basis set has been
used. All ground state geometries were previously optimized at the
corresponding level of theory. Some experimental findings taken from
the literature are also included for comparison \cite{Oddershede1985,Krupenie1972}. The
oscillator strength for each excitation is indicated to the right of the corresponding excitation energy and in the first column of the table the symmetry and
type of the
transition is specified.

\begin{table}[h]
\begin{tabular}{lcccccccccc}
 \multicolumn{1}{c}{Molecule/Trans.}  &  \multicolumn{3}{c}{TD-DFT}
 && \multicolumn{5}{c}{TD-DFTB} & Exp.\\\cline{2-4} \cline{6-10}\\[-0.3cm]
  & $\omega_I$ & $f_I$ &  $\omega_\text{KS}$  & & $\omega_I^\text{new}$ &
$f_I^\text{new}$ &$\omega_I^\text{old}$ &  $f_I^\text{old}$ &  $\omega_\text{KS}$ &\\
\hline
   NO \\
 $~~~\Sigma^+$ ($\pi \rightarrow \pi^*$)     &  6.46 & $<$0.01 & 8.36&& 7.22 & $<$0.01    & 7.49 & $<$0.01& 8.53\\
 $~~~\Pi^{~}$  ($\sigma \rightarrow \pi^*$)  &  6.49 & $<$0.01 & 7.23&& 7.29 & $<$0.01    & 7.77 & 0.00   & 7.80/7.77\\
 $~~~\Delta_{~}$ ($\pi \rightarrow \pi^*$)   &  7.26 & 0.00    & 8.36&& 7.74 & 0.00       & 8.53 & 0.00   & 8.64/8.53\\
 $~~~\Sigma^-$ ($\pi \rightarrow \pi^*$)     &  8.36 & 0.00    & 8.36&& 8.53 & 0.00       & 8.53 & 0.00   & 8.53\\
 $~^2\Pi_{~}$ ($\sigma \rightarrow \pi^*$)   &  8.61 & 0.02    & 7.84&& 8.33 & 0.01       & 7.80 & 0.00   & 7.77/7.80\\
 $~~~\Sigma^-$ ($\pi \rightarrow \pi^*$)     &  8.74 & 0.00    & 8.74&& 8.64 & 0.00       & 8.64 & 0.00   & 8.64\\
 $~^2\Delta_{~}$ ($\pi \rightarrow \pi^*$)   &  9.11 & 0.00    & 8.74&& 9.09 & 0.00       & 8.64 & 0.00   & 8.64\\
 $~~~\Pi_{~}$ ($\sigma^* \rightarrow \pi^*$)   &  11.64 & $<$0.01&12.45&& 12.68 & $<$0.01 & 13.19 & 0.00  & 13.19\\
 $~^2\Sigma^+$ ($\pi \rightarrow \pi^*$)     &  14.00 & 0.35   & 8.47&& 11.90 & 0.63      & 11.65 & 0.50  & 8.64\\
 $~^2\Pi_{~}$ ($\sigma^* \rightarrow \pi^*$)   &  14.82 & 0.38   &12.87&& 14.64 & 0.24    & 13.23 & 0.00  & 13.23\\
\\

 N$_2$ \\

 $~^3\Pi_g$ ($\sigma_g \rightarrow \pi_g$)  &  7.30     &     & 8.20&& 7.48     &   &  8.12    &   & 8.12  &  8.04\\
 $~^3\Sigma_u^+$ ($\pi_u \rightarrow \pi_g$)&  7.42     &     & 9.60&& 6.91     &   &  7.36    &   & 9.01  &  7.75\\
 $~^3\Delta_u$ ($\pi_u \rightarrow \pi_g$)  &  8.24     &     & 9.60&& 7.76     &   &  9.01    &   & 9.01  &  8.88\\
 $~^3\Sigma_u^-$ ($\pi_u \rightarrow \pi_g$)&  9.60     &     & 9.60&& 9.01     &   &  9.01    &   & 9.01  &  9.67\\
 $~^3\Pi_u$ ($\sigma_u \rightarrow \pi_g$)  &  10.37    &     &11.49&& 11.30    &   &  12.06   &   & 12.06 &  11.19\\

 $~^1\Pi_g$ ($\sigma_g \rightarrow \pi_g$)  &  9.05 & 0.00    & 8.20&& 8.71 & 0.00  & 8.12  & 0.00  & 8.12 &  9.31\\
 $~^1\Sigma_u^-$ ($\pi_u \rightarrow \pi_g$)&  9.60 & 0.00    & 9.60&& 9.01 & 0.00  & 9.01  & 0.00  & 9.01 &  9.92\\
 $~^1\Delta_u$ ($\pi_u \rightarrow \pi_g$)  &  10.03 & 0.00   & 9.60&& 9.66 & 0.00  & 9.01  & 0.00  & 9.01 &  10.27\\
 $~^1\Pi_u$ ($\sigma_u \rightarrow \pi_g$)  &  13.53 & 0.42   &11.49&& 13.82 & 0.33 & 12.06 & 0.00  & 12.06&  12.78\\
 $~^1\Sigma_u^+$ ($\pi_u \rightarrow \pi_g$)&  14.84 & 0.77   & 9.60&& 13.02 & 0.98 & 12.75 & 0.80  & 9.01 &  12.96\\

\\
O$_2$ \\

 $~^3\Delta_u$ ($\pi_u \rightarrow \pi_g$)  &  6.39 & 0.00   &  6.90  & &  6.21 & 0.00        & 6.35 & 0.00       & 6.35 &  6.0-6.2\\
 $~^3\Sigma_u^-$ ($\pi_u \rightarrow \pi_g$)&  6.90 & 0.00   &  6.90  & &  6.35 & 0.00        & 6.35 & 0.00       & 6.35 &  6.3-6.5 \\
 $~^3\Pi_g$ ($\sigma_g \rightarrow \pi_g$)  &  7.84 & 0.00   &  7.91  & &  6.80 & 0.00        & 6.79 & 0.00       & 6.79 \\
 $~^3\Sigma_u^+$ ($\pi_u \rightarrow \pi_g$)&  9.00 & 0.18   &  6.90  & &  8.36 & 0.32        & 8.21 & 0.24       & 6.35 &  $\sim$8.6\\
 $~^3\Pi_u$ ($\sigma_u \rightarrow \pi_g$)  &  14.85 & 0.18  & 14.16  & & 15.35 & 0.18       & 14.52 & 0.00      & 14.52 \\

\end{tabular}
\caption{Comparison of vertical excitation energies $\omega_I$ and
  oscillator strengths $f_I$ for TD-DFT with TZP basis set, the
  traditional TD-DFTB method (old) and TD-DFTB with onsite correction
  (new). The PBE functional is used throughout. $\omega_\text{KS}$
  denotes the KS orbital energy difference corresponding to the
  most dominant single particle transition in the many body
  wavefunction, as discussed by Casida in Ref.
  \cite{Casida1995}. Experimental data for N$_2$ and O$_2$ were taken from Ref. \cite{Oddershede1985} and inferred from the potential energy curves in Ref. \cite{Krupenie1972}, respectively. Oscillator strengths are only provided for excitations that are not trivially spin-forbidden. All energies are expressed in eV.}
\label{diatomic}
\end{table}

These molecules have as a common feature that they all possess
low-lying $\sigma \rightarrow \pi^*$ excitations playing an important
role in their absorption spectra. As stated above, a wrong description
of these transitions is a known issue in current TD-DFTB. Below, we
identify yet another failure in the description of some $\pi
\rightarrow \pi^*$ transitions of these compounds.

\begin{figure}[ht]
\begin{center}
\includegraphics[width=0.7\linewidth,angle=0]{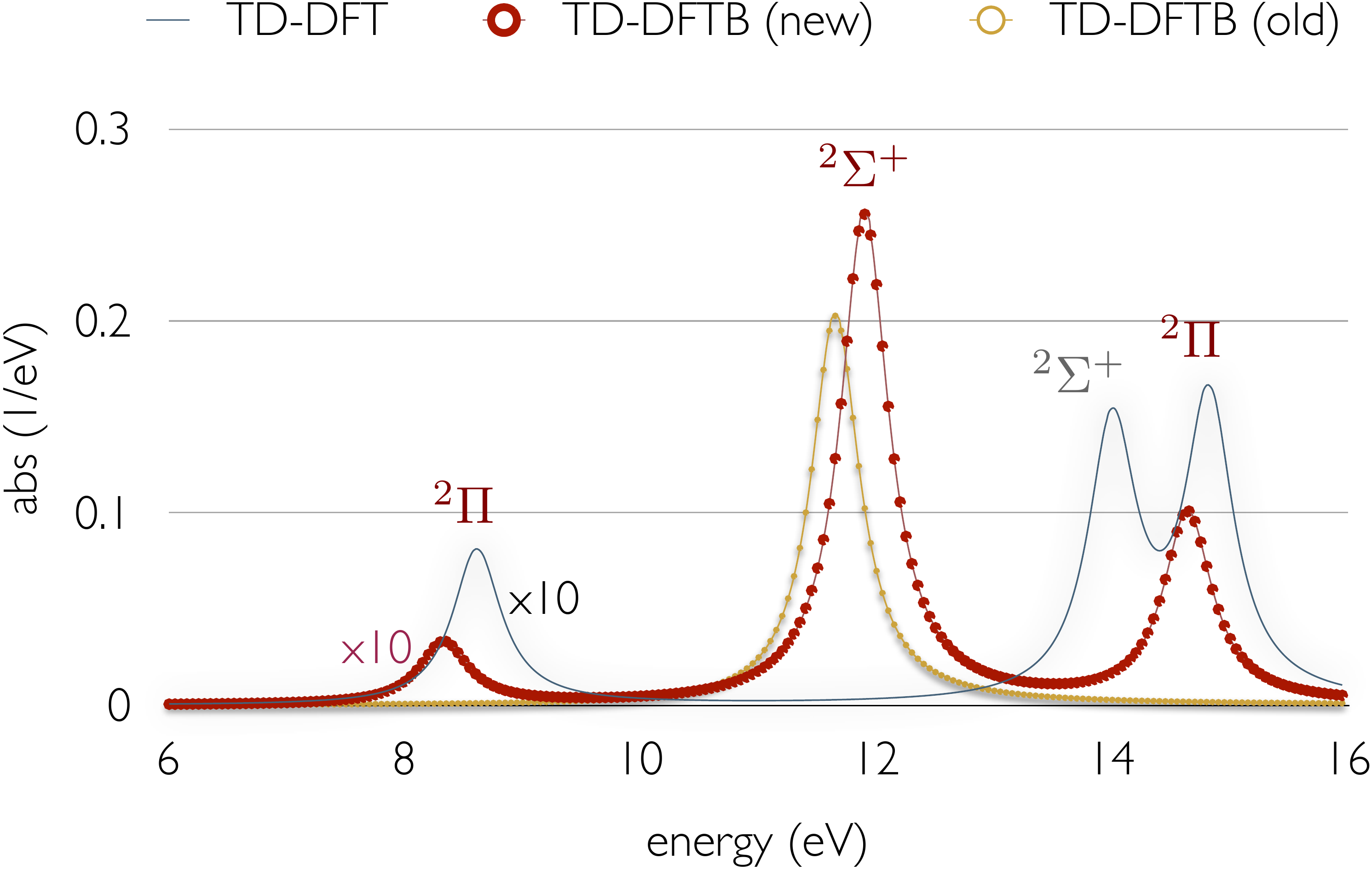}
\caption{\label{fig_NO} Absorption spectrum of nitric oxide as
  obtained with full TD-DFT (PBE/TZP), traditional TD-DFTB (old) and
  TD-DFTB with on-site corrections (new).}
\end{center}
\end{figure}

NO belongs to the symmetry point group C$_{\infty v}$ for which $\Pi$
and $\Sigma^+$ transitions are electric dipole allowed. However,
TD-DFTB describes the former as forbidden. This is due to the
mentioned vanishing of the corresponding transition charge which leads
to an equality of the KS energy difference $\omega_\text{KS}$
(termed   $\omega_{jb\tau}$ in \ref{ABC}) and the excited state
energy $\omega_I$. Within the refined
formulation this failure is successfully overcome as shown in
\ref{diatomic}.
This improvement is specially important for the second
and fourth $\Pi$ transitions with oscillator strengths of 0.01 and
0.24, respectively, which are in agreement with the TD-DFT values of
0.02 and 0.38. Our correction is in this case, essential for providing a correct absorption spectrum (see \ref{fig_NO}) where traditional TD-DFTB is able to describe only the $\Sigma^+$ peak. 

The onsite correction also improves  the
description of $\Pi$ transitions quantitatively. For example, the first $\Pi$
excitation energy is clearly overestimated with respect to
first-principle results. When using our correction the overestimation
is reduced by almost 0.5 eV. Oppositely, the second $\Pi$ excitation
energy is strongly underestimated within traditional TD-DFTB whereas
the corrected calculations return a value (8.33 eV) close to that
from TD-DFT (8.61 eV). The $\Sigma^+$ ($\pi \rightarrow \pi^*$) transitions are also found to be better described within our correction, although the
first (second) $\Sigma^+$ excitation energy is still significantly
overestimated (underestimated). The $\Sigma^-$ ($\pi \rightarrow \pi^*$)
excitations energies are on the other hand unaffected by
our correction. For these transitions no shift from their KS energy differences are seen, which agrees with first-principle observations. More importantly, the onsite correction rectifies the wrong
degeneracy of the transitions $\Sigma^-$ and $\Delta$, which we
identify as another important qualitative issue within traditional
TD-DFTB. The correction also reduces the spin contamination of the doublet $~^2\Pi$ states. 

The electric-dipole-allowed transitions $~^1\Pi_u$ and $~^3\Pi_u$ of
the homonuclear molecules N$_2$ and O$_2$ (point group D$_{\infty
h}$), respectively, are neither correctly described by traditional
TD-DFTB as shown in \ref{diatomic}. When applying the onsite
correction, these excitations become allowed with oscillator strengths
in agreement with ab-initio results. However, it is necessary to
indicate that the excitation energy of the latter transition is somewhat
overestimated, being in better agreement within the non-corrected
formalism. In contrast, the onsite correction greatly improves the agreement of the excitation
energy of N$_2~^1\Pi_u$ with the TD-DFT result. This transition
and the $~^3\Pi_u$ state are degenerate according to traditional results whereas
such degeneracy is broken at the corrected TD-DFTB level. This
degeneracy breaking is in total correspondence with the results
obtained at a higher level of theory.

The original formalism also predicts the degeneracy of the singlet and
triplet $\Sigma^-_u$ and $\Delta_u$ states of N$_2$ with excitation
energy equal to 9.01 eV. This is however only partially confirmed by
the ab-initio results and in total disagreement with the experimental
findings. According to TD-DFT, the triplet and singlet $\Sigma_u^-$
states are degenerate with corresponding excitation energy of 9.60 eV
but the triplet and singlet $\Delta_u$ degeneracy is not observed. On
the other hand, experiments report nearly degenerate $\Sigma_u^-$
states with excitation energies of 9.67 and 9.92 eV for the triplet
and singlet states, respectively, in contrast with TD-DFT findings. This
apparent failure of TD-DFT has been also reproduced elsewhere. \cite{Gross_2000,Bauernschmitt1996}. In a recent letter it was shown that excited states like N$_2$ $\Sigma_u^-$ cannot be described by linear-response TD-DFT as their corresponding excitation energies do not correspond to poles of the response function \cite{Hesselmann2009}. TD-DFTB as an approximation to TD-DFT
unavoidably inherits this issue and our correction is unable to fix
it. However, it does break the wrong degeneracy of the triplet and
singlet $\Delta_u$ states. For O$_2$, a similar degeneracy breaking of
the transitions $\Sigma_u^-$ and $\Delta_u$ within the corrected TD-DFTB can
be observed. This is again in total agreement with first-principle
results.

\subsection{General benchmark}
To assess the general performance of the corrected TD-DFTB method in
terms of vertical excitation energies we have chosen a large benchmark
set defined elsewhere \cite{Schreiber2008} and recently used for the
validation of TD-DFTB \cite{Trani2011}. This set covers $\pi \rightarrow \pi^*$, $n \rightarrow \pi^*$ and $\sigma \rightarrow \pi^*$ excitations for 28
organic compounds comprising unsaturated aliphatic hydrocarbons (group 1), aromatic hydrocarbons and heterocycles (group 2), carbonyl
compounds (group 3) and nucleobases (group 4), and intends to embrace the
most important chromophores in organic photochemistry. For comparison
we have calculated singlet and triplet vertical excitation energies at
the TD-DFT (PBE, TZP) level. To disentangle the
accuracy of the TD-DFTB approximation itself from the quality of
DFTB ground state geometries, the same optimized structures were
used along the benchmark. These geometries were previously obtained at
a MP2 level \cite{Schreiber2008}. In the Supporting Information, an
extended table containing all our calculation results can be
consulted. For further comparison, we also provide the theoretical
best estimates (TBE) for this benchmark set, calculated at the CASPT2
level and reported by Schreiber and co-workers.\cite{Schreiber2008}. Reported experimental values for some vertical excitation energies are additionally given. We also include the KS
energy difference of the corresponding dominant single-particle
transition, which is useful for the analysis of the relative
displacement of the excitation energies with respect to the KS energy
difference as compared to ab-initio TD-DFT.

In \ref{statistics_S} and \ref{statistics_T}, we present a statistical analysis of the collected data for the groups of compounds previously mentioned, for singlet-to-singlet and singlet-to-triplet transitions, respectively. In \ref{statistics_Total} the global statistics for the benchmark set is summarized. Here the analysis has been also split into triplet and singlet excitations. The mean signed difference (MSD) as well as the root-mean-square deviation (RMS) of the TD-DFTB excitation energies with respect to the experiment, the TBE and TD-DFT are reported. Statistics on ab-initio TD-DFT results are also provided to indicate its degree of correspondence with higher level of theory and experiment.

\begin{table}[h]
\begin{tabular}{llcccc}
 \multicolumn{2}{c}{ } & count & TD-DFT &  \multicolumn{2}{c}{TD-DFTB}
 \\\cline{5-6} \\[-0.3cm]
  & & & & $new$ & $old$ \\
\hline
\multicolumn{2}{c}{Group 1} \\

MSD  &  (TD-DFT)    &  13  &       &  0.15 &  0.07 \\
     &  (TBE)       &  13  & -0.52 & -0.37 & -0.45 \\
     &  (Exp.)      &  12  & -0.35 & -0.20 & -0.26 \\

RMS  &  (TD-DFT)    &  13  &       &  0.28 &  0.25 \\
     &  (TBE)       &  13  &  0.62 &  0.47 &  0.53 \\
     &  (Exp.)      &  12  &  0.56 &  0.39 &  0.46 \\
\hline
\multicolumn{2}{c}{Group 2} \\

MSD &  (TD-DFT)     &  53    &       & 0.23 &  0.12 \\
     &  (TBE)       &  53    & -0.35 &-0.12 & -0.23 \\
     &  (Exp.)      &  42    & -0.06 & 0.13 &  0.02 \\

RMS  &  (TD-DFT)    &  53    &       & 0.41 &  0.36 \\
     &  (TBE)       &  53    &  0.54 & 0.37 &  0.43 \\
     &  (Exp.)      &  42    &  0.39 & 0.35 &  0.34 \\
\hline
\multicolumn{2}{c}{Group 3} \\

MSD  &  (TD-DFT)   &   19    &      &  0.56 &  0.37 \\
     &  (TBE)      &   19    &-0.81 & -0.25 & -0.44 \\
     &  (Exp.)     &   13    &-0.67 &  0.07 & -0.08 \\

RMS  &  (TD-DFT)   &   19    &      &  1.06 &  0.93 \\
     &  (TBE)      &   19    & 0.96 &  0.88 &  0.88 \\
     &  (Exp.)     &   13    & 0.88 &  0.70 &  0.64 \\
\hline
\multicolumn{2}{c}{Group 4} \\

MSD  &  (TD-DFT)    &  19    &       &  0.09 &  0.03  \\
     &  (TBE)       &  19    & -0.84 & -0.75 & -0.81  \\
     &  (Exp.)      &  13    & -0.51 & -0.37 & -0.29  \\

RMS  &  (TD-DFT)    &  19    &       &  0.32 &  0.30  \\
     &  (TBE)       &  19    &  0.89 &  0.91 &  0.96  \\
     &  (Exp.)      &  13    &  0.63 &  0.61 &  0.64  \\
\end{tabular}
\caption{Mean signed difference (MSD) and root-mean-square deviation (RMS) of singlet-singlet vertical excitation energies (in eV) calculated within TD-DFTB and TD-DFT (along the rows) with respect to TD-DFT, TBE and experiment (along the columns) where possible.}
\label{statistics_S}
\end{table}

\begin{table}[h]
\begin{tabular}{llcccc}
 \multicolumn{2}{c}{ } & count & TD-DFT &  \multicolumn{2}{c}{TD-DFTB}
 \\\cline{5-6} \\[-0.3cm]
  & & & & $new$ & $old$ \\
\hline
\multicolumn{2}{c}{Group 1} \\

MSD  &  (TD-DFT)    &   13   &       &  0.21 &  0.71 \\
     &  (TBE)       &   13   & -0.32 & -0.11 &  0.39 \\
     &  (Exp.)      &   11   & -0.21 &  0.00 &  0.54 \\

RMS  &  (TD-DFT)    &   13   &       &  0.32 &  0.78 \\
     &  (TBE)       &   13   &  0.38 &  0.18 &  0.48 \\
     &  (Exp.)      &   11   &  0.25 &  0.21 &  0.61 \\
\hline
\multicolumn{2}{c}{Group 2} \\

MSD  &  (TD-DFT)    &   36   &       &  0.28 &  0.55 \\
     &  (TBE)       &   36   & -0.52 & -0.24 &  0.03 \\
     &  (Exp.)      &   12   & -0.19 &  0.26 &  0.59 \\

RMS  &  (TD-DFT)    &   36   &       &  0.41 &  0.63 \\
     &  (TBE)       &   36   &  0.63 &  0.41 &  0.45 \\
     &  (Exp.)      &   12   &  0.33 &  0.38 &  0.69 \\
\hline
\multicolumn{2}{c}{Group 3} \\

MSD  &  (TD-DFT)    &  14    &       &  0.42 &  0.79 \\
     &  (TBE)       &  14    & -0.61 & -0.20 &  0.17 \\
     &  (Exp.)      &   7    & -0.53 & -0.13 &  0.27 \\

RMS  &  (TD-DFT)    &  14    &       &  0.47 &  0.83 \\
     &  (TBE)       &  14    &  0.67 &  0.41 &  0.47 \\
     &  (Exp.)      &   7    &  0.57 &  0.39 &  0.57 \\
\end{tabular}
\caption{Mean signed difference (MSD) and root-mean-square deviation (RMS) of singlet-triplet vertical excitation energies (in eV) calculated within TD-DFTB and TD-DFT (along the rows) with respect to TD-DFT, TBE and experiment (along the columns) where possible.}
\label{statistics_T}
\end{table}

\begin{table}[h]
\begin{tabular}{llcccc}
 \multicolumn{2}{c}{ } & count & TD-DFT &  \multicolumn{2}{c}{TD-DFTB}
 \\\cline{5-6} \\[-0.3cm]
  & & & & $new$ & $old$ \\
\hline
\multicolumn{2}{c}{Singlets} \\

MSD  &  (TD-DFT)    &  104    &       &  0.25 &  0.14 \\
     &  (TBE)       &  104    & -0.55 & -0.29 & -0.40 \\
     &  (Exp.)      &   80    & -0.28 & -0.01 & -0.12 \\

RMS  &  (TD-DFT)    &  104    &       &  0.57 &  0.50 \\
     &  (TBE)       &  104    &  0.71 &  0.63 &  0.66 \\
     &  (Exp.)      &   80    &  0.56 &  0.48 &  0.47 \\
\hline
\multicolumn{2}{c}{Triplets} \\

MSD  &  (TD-DFT)    &   63   &       &  0.29 &  0.63 \\
     &  (TBE)       &   63   & -0.50 & -0.20 &  0.14 \\
     &  (Exp.)      &   30   & -0.27 &  0.07 &  0.50 \\

RMS  &  (TD-DFT)    &   63   &       &  0.40 &  0.71 \\
     &  (TBE)       &   63   &  0.59 &  0.38 &  0.46 \\
     &  (Exp.)      &   30   &  0.38 &  0.33 &  0.63 \\
\hline
\multicolumn{2}{c}{Total} \\

MSD  &  (TD-DFT)    &   167   &       &  0.27 &  0.33 \\
     &  (TBE)       &   167   & -0.53 & -0.26 & -0.20 \\
     &  (Exp.)      &   110   & -0.28 &  0.01 &  0.05 \\

RMS  &  (TD-DFT)    &   167   &       &  0.51 &  0.59 \\
     &  (TBE)       &   167   &  0.67 &  0.55 &  0.59 \\
     &  (Exp.)      &   110   &  0.52 &  0.44 &  0.52 \\
\end{tabular}
\caption{Global statistics for vertical excitation energies (in eV) calculated within TD-DFTB and TD-DFT (along the rows) with respect to TD-DFT, TBE and experiment (along the columns) where possible. }
\label{statistics_Total}
\end{table}

To assess the validity of our approximation, we focus on the
comparison with the TD-DFT (PBE) data set. It is important to recall
that TD-DFTB parameters are calculated at this level of theory and the
main aim of our approach is to improve its agreement with respect to
the TD-DFT description. Since TD-DFT at the PBE level is of course an
approximation itself, we are also interested in examining the accuracy
of our method compared to a higher level of theory and experiment. It
is worth mentioning that the subset of singlet-triplet excitations for
which experimental observations are provided is around 50\% smaller
than the original set, but we consider it still suitable to perform a
statistical analysis. The TBEs are available for the complete
benchmark set on the other hand and they are fairly close to their corresponding
experimental values, with a RMS error of 0.24 eV for triplet states
and 0.38 eV for singlet states. On average, both singlet and triplet
TBEs are slightly overestimated (MSD = 0.15 eV and 0.20 eV,
respectively) which should be taken into account in the further
analysis. Consistent with earlier studies,\cite{Jacquemin2009,Jacquemin2010}  TD-DFT excitation energies at the PBE level
appear to be somewhat underestimated with regard to experimental
findings as a general trend (MSD = -0.28 eV and -0.27 eV for singlet and triplet states, respectively).

It can bee seen from \ref{statistics_Total} that with respect to
TD-DFT, traditional TD-DFTB performs better for singlet-singlet than
for singlet-triplet excitations. It should be  pointed out however,
that triplet states are in better agreement with the TBE values as it
has been previously noticed \cite{Trani2011}. Nevertheless, compared to the
collected experimental findings, singlet states also appear to be
better described. In fact, \ref{statistics_Total} shows a significant
overestimation of triplet state energies compared to experiment (MSD =
0.50 eV) within traditional TD-DFTB. One of the main effects of our
correction is the significant improvement of singlet-triplet
excitation energies taking TD-DFT and experimental values as a
reference. Within the refined formulation, the RMS error is reduced by
0.3 eV with respect to both TD-DFT and experiment. More importantly, the RMS error with respect to the
latter (0.33 eV) is slightly lower than that for TD-DFT (0.38 eV). A similar accuracy for both TD-DFT and refined TD-DFTB can be seen for the first two group of compounds whereas for the third group, the agreement with experimental data is somewhat better for our method (see \ref{statistics_T}). In addition, whereas \ref{statistics_Total} still indicates some overestimation of the corrected TD-DFTB results with respect to TD-DFT, the MSD of our refined approach compared to experiment becomes very small, which shows that the corrected excitation energies are scattered around the experimental values. Specifically, in the group of aromatic hydrocarbons and heterocycles, excitation energies are overestimated whereas there is some underestimation for the carbonyl compounds. Only with regard to TBE, the refined excitation energies appear to be underestimated for every group of compounds.

In contrast to the observations for singlet-triplet transitions, the
traditional TD-DFTB method is,
on average, in slightly better agreement with TD-DFT. The RMS deviations of the corrected and non-corrected methods
from TD-DFT are 0.57 eV and 0.50 eV, respectively. On the other hand,
regarding the experimental references and the TBEs, both approaches perform with similar accuracy. The refined formalism returns
overestimated excitation energies according to TD-DFT results,
although with respect to experiment they are again spread around the
reference values. In this case, the overestimation for the second group of compounds is compensated with the
underestimation for the aliphatic hydrocarbons and the nucleobases, thus leading to an almost vanishing MSD (-0.01
eV). The results for the latter mentioned group of compounds are the least overestimated with respect to
TD-DFT, with a MSD of only 0.09 eV. By contrast, the underestimation with respect to experiment is significant (MSD = -0.37 eV) and increases even more by taking the
TBEs as the reference, for which the MSD amounts to -0.75 eV. For this
group, the RMS error of TD-DFTB with respect to TBEs is remarkably
high (0.91 eV and 0.96 eV for the corrected and non-corrected
approaches, respectively). The limited agreement between the TD-DFTB
excitation energies of the nucleobases and their TBE counterparts has
been already pointed out by Trani and co-workers.\cite{Trani2011}
However, it is necessary to notice that this failure should be rather
attributed to TD-PBE itself and not to TD-DFTB as an approximation. In
fact, the worst agreement between TD-DFT and TBE along the benchmark
set is found for the carbonyl compounds and the nucleobases, with RMS
errors of 0.96 eV and 0.89 eV, respectively. A very recent study by
Foster and Wong indeed shows that conventional
semi-local functionals fail in the description of the optical
properties of nucleobases, while tuned range-separated functionals
offer significant improvements.\cite{Foster2012}

The overall analysis (see bottom of \ref{statistics_Total}) leads to somewhat smaller RMS errors for corrected TD-DFTB compared to the non-corrected formalism. It
should be indicated, that despite the important improvements for the
triplets states within the corrected formulation, the benchmark set
for singlet-singlet excitations is comparatively larger, conceding
more importance to these kind of transitions within the total
balance. As a general behavior, it can be stated that TD-DFTB singlet
excited states are shifted up in energy when the onsite correction is
switched on. Inversely, our correction tends to shift triplet states
down. This effect is particularly noticeable for $n \rightarrow \pi^*$
and $\sigma \rightarrow \pi^*$ excitations. The excitation energies
for these transitions are either identically equal or (for few cases)
very close to their corresponding KS energy differences at the
non-corrected TD-DFTB level, resulting in degenerate or
nearly-degenerate singlet and triplet states (see Supporting
Information). The only exception for this statement are the triplet $B_{1u}$
states of tetrazine, for which an energy displacement occurs also for
traditional TD-DFTB, although clearly underestimated with respect to
the TD-DFT results. Within the onsite correction the excitation
energies are either shifted down for triplets or shifted up for
singlets with respect to the KS energy difference, leading to a
singlet-triplet gap in accordance with the observations at the TD-DFT level. A similar degeneracy breaking was shown earlier for N$_2$.

Considering only $n \rightarrow \pi^*$ and $\sigma \rightarrow \pi^*$
transitions, the RMS error of singlet-triplet excitation energies at the
corrected (non-corrected) TD-DFTB level as compared to TD-DFT is 0.58
(0.82) eV. These kind of excitations are evidently difficult
cases for the traditional formalism and an important improvement is obtained with the onsite correction, although the major quantitative enhancement of the latter approach is for singlet-triplet $\pi\rightarrow\pi^*$ transitions. Indeed, $n \rightarrow \pi^*$ and $\sigma \rightarrow \pi^*$ excitation energies are still strongly overestimated at the corrected TD-DFTB level with respect to TD-DFT, with MSD of 0.49 and 0.32 eV for triplet and singlet states, respectively. However, if we compare our findings with the TBEs, those are by contrast, significantly underestimated (MSD = -0.21 and -0.40 eV for singlet-triplet and singlet-singlet transitions, respectively) and the RMS deviations for both corrected and non-corrected formalisms are similar.

\begin{figure}[ht]
\begin{center}
\includegraphics[width=0.7\linewidth,angle=0]{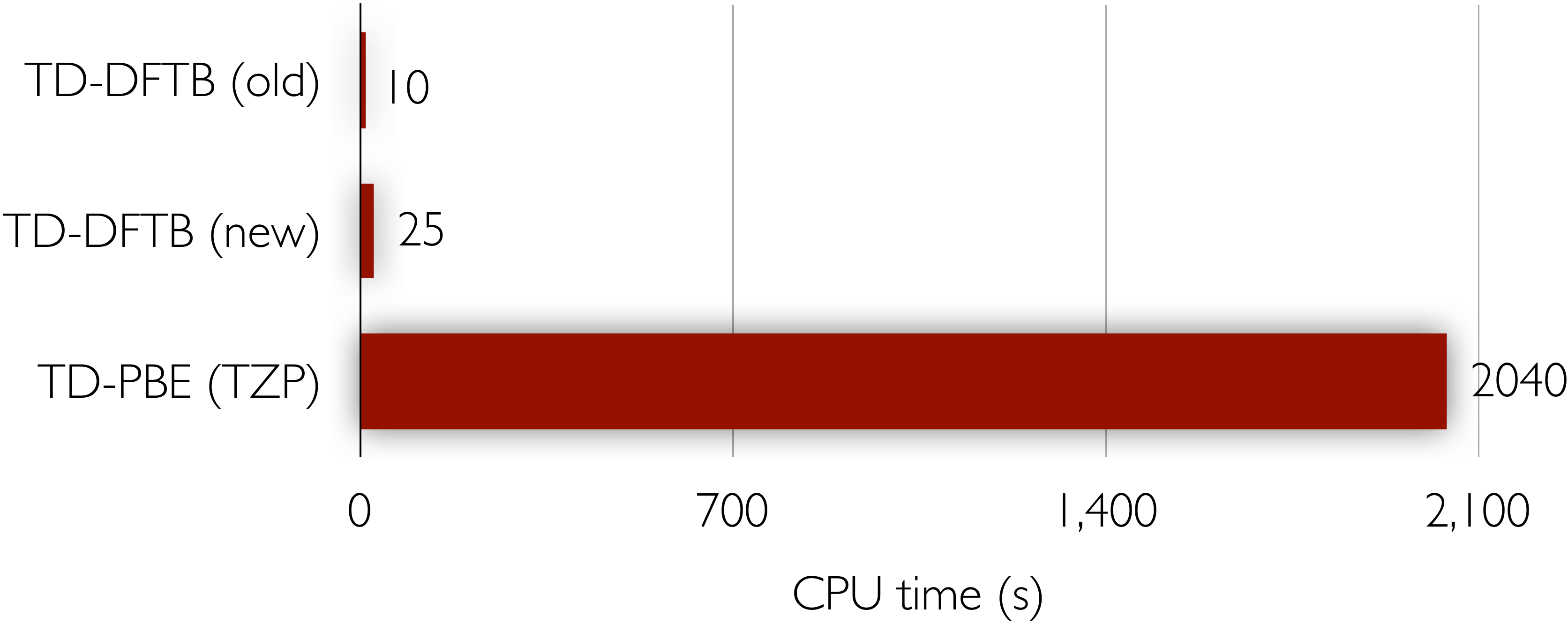}
\caption{\label{fig_cpuTime} Total computational time required for calculating the low-lying excitation energies of a set of 28 molecules with traditional TD-DFTB (old), refined TD-DFTB (new) and full TD-DFT (PBE/TZP).}
\end{center}
\end{figure}

\section{CONCLUSIONS}

We have introduced and implemented new extensions of the
time-dependent density functional based tight binding (TD-DFTB)
method. By taking the formalism beyond a mere Mulliken approximation,
we have been able to surpass the wrong description of $n \rightarrow
\pi^*$ and $\sigma \rightarrow \pi^*$ excitations within TD-DFTB. It was
shown that particularly for the molecules O$_2$, NO and N$_2$
these kind of transitions play an important role in the low energy
optical spectra. Our refined formulation is in this case essential to
obtain a qualitatively correct spectrum. Moreover, for the close-shell
molecule N$_2$, the wrong $\sigma \rightarrow \pi^*$ triplet-singlet
degeneracy encountered by using the original approach was successfully
overcome. We additionally identified the erroneous degeneracy of $\pi
\rightarrow \pi^*$ excitations with irreducible representations $\Sigma^-$ and $\Delta$ for
the investigated diatomic molecules  as another failure of traditional
TD-DFTB. The new formulation now delivers the correct behavior as
compared to TD-DFT observations. Along with the qualitative
improvement, a better numerical agreement with TD-DFT results can also
be perceived for these systems.

To extend the analysis of the performance of our correction, we
calculated the low-lying vertical excitation energies for a set of 28
benchmark compounds. We found a considerable improvement in the
agreement with TD-DFT and experiment in terms of singlet-triplet
excitation energies, whereas singlet-singlet energies have the same
overall quality as in the traditional scheme. The accuracy of the corrected formalism is similar to that of TD-DFT for both triplet and singlet states, whereas the computational time of the former method is roughly eighty times smaller (see \ref{fig_cpuTime}).

\section{ACKNOWLEDGEMENT}
Financial support from the Deutsche Forschungsgemeinschaft (GRK 1570
and SPP 1243) and the DAAD is gratefully acknowledged.

\appendix
\section{Onsite correction to the total energy}
\label{App_ener}

Let $\{\phi_{\alpha}\}$ and $\{\phi_{\beta}\}$ be a complete set of
real orbitals placed at atom $A$ and $B$,
respectively. The orbitals are assumed to be orthonormal within each
set, while the overlap between orbitals on different atoms is given by
$S_{\alpha\beta}$. Let also $\mu \in A$, $\nu \in B$, $\kappa \in C$ and $\lambda \in D$ unless otherwise specified. Then, the orbitals $\phi_{\mu}$ and $\phi_{\nu}$ can be expanded as

\begin{equation}
 \phi_{\mu}(\mathbf{r}) = \sum_{\beta \in B} S_{\beta\mu} \phi_{\beta}(\mathbf{r}), ~~~ \phi_{\nu}(\mathbf{r}) = \sum_{\alpha \in A} S_{\alpha\nu} \phi_{\alpha}(\mathbf{r}).
\end{equation}
Hence, the orbital product can be expressed as

\begin{equation}
 \phi_{\mu}(\mathbf{r})\phi_{\nu}(\mathbf{r}) = \frac{1}{2} \left( \sum_{\alpha \in A} S_{\alpha\nu} \phi_{\alpha}(\mathbf{r})\phi_{\mu}(\mathbf{r}) + \sum_{\beta \in B} S_{\beta\mu} \phi_{\beta}(\mathbf{r})\phi_{\nu}(\mathbf{r}) \right),
\end{equation}
and more conveniently as

\begin{equation}
 \phi_{\mu}(\mathbf{r})\phi_{\nu}(\mathbf{r}) =\frac{1}{2} S_{\mu\nu} \left( |\phi_{\mu}(\mathbf{r})|^2 +|\phi_{\nu}(\mathbf{r})|^2 \right) + \frac{1}{2} \left( \sum_{\alpha \in A}^{\alpha \neq \mu} S_{\alpha\nu} \phi_{\alpha}(\mathbf{r})\phi_{\mu}(\mathbf{r}) + \sum_{\beta \in B}^{\beta\neq\nu} S_{\beta\mu} \phi_{\beta}(\mathbf{r})\phi_{\nu}(\mathbf{r}) \right),
\label{AO_product}
\end{equation}
where the first term accounts for the Mulliken approach. Let us now denote with $ (\mu\nu|\kappa\lambda)$ a two-electron integral with an arbitrary kernel. Using \ref{AO_product}, the integrals $ (\mu\nu|\kappa\lambda)$ can then be expanded as

\begin{eqnarray}
 (\mu\nu|\kappa\lambda) &=& \frac{1}{4}S_{\mu\nu} S_{\kappa\lambda} \left[(\mu\mu|\kappa\kappa) + (\mu\mu|\lambda\lambda) + (\nu\nu|\kappa\kappa) + (\nu\nu|\lambda\lambda)\right] \nonumber \\
                        &+& \frac{1}{4}S_{\mu\nu} \left( \sum_{\gamma \in C}^{\gamma\neq\kappa} S_{\gamma\lambda} \left[(\mu\mu|\kappa\gamma) + (\nu\nu|\kappa\gamma)\right] + \sum_{\delta \in D}^{\delta\neq\lambda} S_{\delta\kappa} \left[(\mu\mu|\delta\lambda) + (\nu\nu|\delta\lambda)\right] \right)  \nonumber \\
                        &+& \frac{1}{4}S_{\kappa\lambda} \left( \sum_{\alpha \in A}^{\alpha\neq\mu} S_{\alpha\nu} \left[(\mu\alpha|\kappa\kappa) + (\mu\alpha|\lambda\lambda)\right] + \sum_{\beta \in B}^{\beta\neq\nu} S_{\beta\mu} \left[(\beta\nu|\kappa\kappa) + (\beta\nu|\lambda\lambda)\right] \right)  \nonumber \\
                        &+& \frac{1}{4} \left( \sum_{\alpha \in A}^{\alpha\neq\mu} \sum_{\gamma \in C}^{\gamma\neq\kappa} S_{\alpha\nu}S_{\gamma\lambda} (\mu\alpha|\kappa\gamma) + \sum_{\alpha \in A}^{\alpha\neq\mu}\sum_{\delta \in D}^{\delta\neq\lambda} S_{\alpha\nu}S_{\delta\kappa} (\mu\alpha|\delta\lambda) \right) \nonumber \\
                        &+& \frac{1}{4} \left( \sum_{\beta \in B}^{\beta\neq\nu}\sum_{\gamma \in C}^{\gamma\neq\kappa} S_{\beta\mu}S_{\gamma\lambda} (\beta\nu|\kappa\gamma) + \sum_{\beta \in B}^{\beta\neq\nu}\sum_{\delta \in D}^{\delta\neq\lambda} S_{\beta\mu}S_{\delta\kappa}(\beta\nu|\delta\lambda) \right).
\label{int_exp}
\end{eqnarray}

Note that the expression \ref{int_exp} is exact as long as the
AO sets  $\{\phi_{\alpha}\}$, $\{\phi_{\beta}\}$, $\{\phi_{\gamma}\}$
and  $\{\phi_{\delta}\}$ are complete. The first line in eq
\ref{int_exp} contains the leading terms, which include
Coulomb-like integrals. Truncation of the expansion up to the first
line accounts for the Mulliken approach. A next level of approximation
demands the inclusion of fully-onsite exchange-like integrals, i.e. $(\mu\nu|\mu\nu)$, with $\mu,\nu \in A$ and $\mu \neq \nu$. At this level of theory the general four-center integrals can be expressed as

\begin{equation}
 (\mu\nu|\kappa\lambda) \approx (\mu\nu|\kappa\lambda)_\text{mull} + (\mu\nu|\kappa\lambda)_\text{ons},
\end{equation}
where

\begin{equation}
 (\mu\nu|\kappa\lambda)_\text{mull} = \frac{1}{4}S_{\mu\nu} S_{\kappa\lambda} \left[(\mu\mu|\kappa\kappa) + (\mu\mu|\lambda\lambda) + (\nu\nu|\kappa\kappa) + (\nu\nu|\lambda\lambda)\right],
\end{equation}
and

\begin{eqnarray}
  (\mu\nu|\kappa\lambda)_\text{ons} &=& \frac{1}{4} \sum_{\alpha \in A}^{\alpha\neq\mu} S_{\alpha\nu}S_{\alpha\lambda} (\mu\alpha|\mu\alpha) \delta_{\mu\kappa} + \frac{1}{4}\sum_{\alpha \in A}^{\alpha\neq\mu} S_{\alpha\nu} S_{\alpha\kappa} (\mu\alpha|\mu\alpha)                      \delta_{\mu\lambda} \nonumber \\
                               &+& \frac{1}{4} \sum_{\beta \in B}^{\beta\neq\nu} S_{\beta\mu}S_{\beta\lambda} (\beta\nu|\beta\nu) \delta_{\nu\kappa} + \frac{1}{4}\sum_{\beta \in B}^{\beta\neq\nu} S_{\beta\mu} S_{\beta\kappa} (\beta\nu|\beta\nu)       \delta_{\nu\lambda} \nonumber \\
                               &+& \frac{1}{4} S_{\kappa\nu}S_{\mu\lambda} (\mu\kappa|\mu\kappa) \delta_{AC} (1-\delta_{\mu\kappa}) + \frac{1}{4} S_{\lambda\nu}S_{\mu\kappa} (\mu\lambda|\mu\lambda) \delta_{AD} (1-\delta_{\mu\lambda}) \nonumber \\
                               &+& \frac{1}{4} S_{\kappa\mu}S_{\nu\lambda} (\nu\kappa|\nu\kappa) \delta_{BC} (1-\delta_{\nu\kappa}) + \frac{1}{4} S_{\lambda\mu}S_{\nu\kappa} (\nu\lambda|\nu\lambda) \delta_{BD} (1-\delta_{\nu\lambda})\,.
\label{ons_int}
\end{eqnarray}

The contribution to the DFTB total energy originating from this correction is written as

\begin{equation}
 E_\text{ons} = \frac{1}{2} \sum_{\sigma\tau} \sum_{\mu\nu\kappa\lambda} \Delta P_{\mu\nu}^{\sigma} \left(\mu \nu | f^\text{hxc}_{\rho_\sigma \rho_\tau}[\rho_0]| \kappa \lambda \right)_\text{ons}\,\Delta P_{\kappa\lambda}^{\tau}.
\label{Eons}
\end{equation}
After substituting \ref{ons_int} in \ref{Eons}, we
finally arrive at 

\begin{equation}
 E_\text{ons} = \sum_{\sigma\tau} \sum_{A} \sum_{\mu\nu\in A}^{\mu
   \neq \nu}  \Delta \tilde{P}_{\mu\nu}^{\sigma}  \left(\mu \nu |
   f^\text{hxc}_{\rho_\sigma \rho_\tau}[\rho_0]| \mu \nu \right) \Delta \tilde{P}_{\mu\nu}^{\tau},
\end{equation}
where $\tilde{P}_{\mu\nu}^{\sigma}$ are the matrix elements of the
dual density matrix defined in \ref{dualP}.

\section{Onsite correction to dipole matrix elements}
\label{App_dip}

To derive an expression for the KS dipole matrix elements, we expand the KS orbitals into a set of localized atom-centered AO, $\psi_{s\sigma} = \sum_{\mu} c_{\mu s}^{\sigma} \phi_{\mu}$. Thus, the dipole matrix elements read

\begin{equation}
  \langle \psi_{s\sigma} | \mathbf{\hat{r}} | \psi_{t\sigma} \rangle =
  \sum_{\mu\nu} c_{\mu s}^{\sigma}  \langle \mu | \mathbf{\hat{r}} |
  \nu \rangle c_{\nu t}^{\sigma}
\label{KS_dip}
\end{equation}

Let $\mu \in A$ and $\nu \in B$ unless otherwise indicated. Using the orbital product expansion \ref{AO_product}, the AO dipole matrix elements, $\langle \mu | \mathbf{\hat{r}} | \nu \rangle$, can be expressed as follows,

\begin{equation}
  \langle \mu | \mathbf{\hat{r}} | \nu \rangle = \frac{1}{2} S_{\mu\nu} \left( \mathbf{R}_{A} + \mathbf{R}_{B} \right) + \frac{1}{2} \left( \sum_{\alpha \in A}^{\alpha \neq \mu} S_{\alpha\nu} \langle \alpha | \mathbf{\hat{r}} | \mu \rangle + \sum_{\beta \in B}^{\beta\neq\nu} S_{\beta\mu} \langle \beta | \mathbf{\hat{r}} | \nu \rangle \right),
\label{AO_dip}
\end{equation}
where  $\mathbf{R}_{A} = \langle \mu | \mathbf{\hat{r}} | \mu \rangle$ and $\mathbf{R}_{B} = \langle \nu | \mathbf{\hat{r}} | \nu \rangle$ denote the positions of centers A and B, respectively. After substituting \ref{AO_dip} in \ref{KS_dip}, we finally have

\begin{equation}
 \langle \psi_{s\sigma} | \mathbf{\hat{r}} | \psi_{t\sigma} \rangle = \sum_{A} \mathbf{R}_{A} q_{A}^{st\sigma} + \sum_A \sum_{\mu\nu \in A}^{\mu \neq \nu} P_{\mu\nu}^{st\sigma}  \langle \mu | \mathbf{\hat{r}} | \nu\rangle,
\end{equation}
where the definition \ref{TDM} was additionally employed.


\bibliography{../../Combined}

\providecommand*\mcitethebibliography{\thebibliography}
\csname @ifundefined\endcsname{endmcitethebibliography}
  {\let\endmcitethebibliography\endthebibliography}{}
\begin{mcitethebibliography}{34}
\providecommand*\natexlab[1]{#1}
\providecommand*\mciteSetBstSublistMode[1]{}
\providecommand*\mciteSetBstMaxWidthForm[2]{}
\providecommand*\mciteBstWouldAddEndPuncttrue
  {\def\EndOfBibitem{\unskip.}}
\providecommand*\mciteBstWouldAddEndPunctfalse
  {\let\EndOfBibitem\relax}
\providecommand*\mciteSetBstMidEndSepPunct[3]{}
\providecommand*\mciteSetBstSublistLabelBeginEnd[3]{}
\providecommand*\EndOfBibitem{}
\mciteSetBstSublistMode{f}
\mciteSetBstMaxWidthForm{subitem}{(\alph{mcitesubitemcount})}
\mciteSetBstSublistLabelBeginEnd
  {\mcitemaxwidthsubitemform\space}
  {\relax}
  {\relax}

\bibitem[Runge and Gross(1984)Runge, and Gross]{runge1984dft}
Runge,~E.; Gross,~E. K.~U. \emph{Phys. Rev. Lett.} \textbf{1984}, \emph{52},
  997--1000\relax
\mciteBstWouldAddEndPuncttrue
\mciteSetBstMidEndSepPunct{\mcitedefaultmidpunct}
{\mcitedefaultendpunct}{\mcitedefaultseppunct}\relax
\EndOfBibitem
\bibitem[Casida and Huix-Rotllant(2012)Casida, and Huix-Rotllant]{Casida2012}
Casida,~M.~E.; Huix-Rotllant,~M. \emph{Annu. Rev. Phys. Chem.} \textbf{2012},
  \emph{63}, 287\relax
\mciteBstWouldAddEndPuncttrue
\mciteSetBstMidEndSepPunct{\mcitedefaultmidpunct}
{\mcitedefaultendpunct}{\mcitedefaultseppunct}\relax
\EndOfBibitem
\bibitem[Niehaus et~al.(2001)Niehaus, Suhai, Della~Sala, Lugli, Elstner,
  Seifert, and Frauenheim]{Niehaus2001a}
Niehaus,~T.~A.; Suhai,~S.; Della~Sala,~F.; Lugli,~P.; Elstner,~M.; Seifert,~G.;
  Frauenheim,~T. \emph{Phys. Rev. B} \textbf{2001}, \emph{63}, 085108\relax
\mciteBstWouldAddEndPuncttrue
\mciteSetBstMidEndSepPunct{\mcitedefaultmidpunct}
{\mcitedefaultendpunct}{\mcitedefaultseppunct}\relax
\EndOfBibitem
\bibitem[Heringer et~al.(2007)Heringer, Niehaus, Wanko, and
  Frauenheim]{heringer2007aes}
Heringer,~D.; Niehaus,~T.~A.; Wanko,~M.; Frauenheim,~T. \emph{J. Comp. Chem.}
  \textbf{2007}, \emph{28}, 2589--601\relax
\mciteBstWouldAddEndPuncttrue
\mciteSetBstMidEndSepPunct{\mcitedefaultmidpunct}
{\mcitedefaultendpunct}{\mcitedefaultseppunct}\relax
\EndOfBibitem
\bibitem[Yam et~al.(2003)Yam, Yokojima, and Chen]{Yam2003}
Yam,~C.~Y.; Yokojima,~S.; Chen,~G.~H. \emph{Physical Review B} \textbf{2003},
  \emph{68}, 153105\relax
\mciteBstWouldAddEndPuncttrue
\mciteSetBstMidEndSepPunct{\mcitedefaultmidpunct}
{\mcitedefaultendpunct}{\mcitedefaultseppunct}\relax
\EndOfBibitem
\bibitem[Niehaus et~al.(2005)Niehaus, Heringer, Torralva, and
  Frauenheim]{Niehaus2005}
Niehaus,~T.~A.; Heringer,~D.; Torralva,~B.; Frauenheim,~T. \emph{Eur. Phys. J.
  D} \textbf{2005}, \emph{35}, 467--477\relax
\mciteBstWouldAddEndPuncttrue
\mciteSetBstMidEndSepPunct{\mcitedefaultmidpunct}
{\mcitedefaultendpunct}{\mcitedefaultseppunct}\relax
\EndOfBibitem
\bibitem[Mitric et~al.(2009)Mitric, Werner, Wohlgemuth, Seifert, and
  Bonacic-Kouteck{\`y}]{Mitric2009}
Mitric,~R.; Werner,~U.; Wohlgemuth,~M.; Seifert,~G.; Bonacic-Kouteck{\`y},~V.
  \emph{J. Phys. Chem. A} \textbf{2009}, \emph{113}, 12700\relax
\mciteBstWouldAddEndPuncttrue
\mciteSetBstMidEndSepPunct{\mcitedefaultmidpunct}
{\mcitedefaultendpunct}{\mcitedefaultseppunct}\relax
\EndOfBibitem
\bibitem[Jakowski et~al.(2012)Jakowski, Irle, and Morokuma]{Jakowski2012}
Jakowski,~J.; Irle,~S.; Morokuma,~K. \emph{Phys. Chem. Chem. Phys.}
  \textbf{2012}, \emph{14}, 6273\relax
\mciteBstWouldAddEndPuncttrue
\mciteSetBstMidEndSepPunct{\mcitedefaultmidpunct}
{\mcitedefaultendpunct}{\mcitedefaultseppunct}\relax
\EndOfBibitem
\bibitem[Wang et~al.(2011)Wang, Yam, Frauenheim, Chen, and Niehaus]{Wang2011}
Wang,~Y.; Yam,~C.-Y.; Frauenheim,~T.; Chen,~G.; Niehaus,~T. \emph{Chem. Phys.}
  \textbf{2011}, \emph{391}, 69, <ce:title>Open problems and new solutions in
  time dependent density functional theory</ce:title>\relax
\mciteBstWouldAddEndPuncttrue
\mciteSetBstMidEndSepPunct{\mcitedefaultmidpunct}
{\mcitedefaultendpunct}{\mcitedefaultseppunct}\relax
\EndOfBibitem
\bibitem[Trani et~al.(2011)Trani, Scalmani, Zheng, Carnimeo, Frisch, and
  Barone]{Trani2011}
Trani,~F.; Scalmani,~G.; Zheng,~G.; Carnimeo,~I.; Frisch,~M.; Barone,~V.
  \emph{J. Chem. Theory Comput.} \textbf{2011}, \emph{7}, 3304\relax
\mciteBstWouldAddEndPuncttrue
\mciteSetBstMidEndSepPunct{\mcitedefaultmidpunct}
{\mcitedefaultendpunct}{\mcitedefaultseppunct}\relax
\EndOfBibitem
\bibitem[Niehaus(2009)]{Niehaus2009}
Niehaus,~T.~A. \emph{J. Mol. Struct.: THEOCHEM} \textbf{2009}, \emph{914},
  38\relax
\mciteBstWouldAddEndPuncttrue
\mciteSetBstMidEndSepPunct{\mcitedefaultmidpunct}
{\mcitedefaultendpunct}{\mcitedefaultseppunct}\relax
\EndOfBibitem
\bibitem[K\"{o}hler et~al.(2001)K\"{o}hler, Seifert, Gerstmann, Elstner,
  Overhof, and Frauenheim]{Kohler2001}
K\"{o}hler,~C.; Seifert,~G.; Gerstmann,~U.; Elstner,~M.; Overhof,~H.;
  Frauenheim,~T. \emph{Phys. Chem. Chem. Phys.} \textbf{2001}, \emph{3},
  5109--5114\relax
\mciteBstWouldAddEndPuncttrue
\mciteSetBstMidEndSepPunct{\mcitedefaultmidpunct}
{\mcitedefaultendpunct}{\mcitedefaultseppunct}\relax
\EndOfBibitem
\bibitem[Frauenheim et~al.(2002)Frauenheim, Seifert, Elstner, Niehaus, Kohler,
  Amkreutz, Sternberg, Hajnal, Di~Carlo, and Suhai]{Frauenheim2002}
Frauenheim,~T.; Seifert,~G.; Elstner,~M.; Niehaus,~T.; Kohler,~C.;
  Amkreutz,~M.; Sternberg,~M.; Hajnal,~Z.; Di~Carlo,~A.; Suhai,~S.
  \emph{Journal Of Physics-Condensed Matter} \textbf{2002}, \emph{14},
  3015--3047\relax
\mciteBstWouldAddEndPuncttrue
\mciteSetBstMidEndSepPunct{\mcitedefaultmidpunct}
{\mcitedefaultendpunct}{\mcitedefaultseppunct}\relax
\EndOfBibitem
\bibitem[Casida(1995)]{Casida1995}
Casida,~M.~E. In \emph{Recent Advances in Density Functional Methods, Part I};
  Chong,~D., Ed.; World Scientific, 1995; Chapter Time-dependent Density
  Functional Response Theory for Molecules, pp 155--192\relax
\mciteBstWouldAddEndPuncttrue
\mciteSetBstMidEndSepPunct{\mcitedefaultmidpunct}
{\mcitedefaultendpunct}{\mcitedefaultseppunct}\relax
\EndOfBibitem
\bibitem[Niehaus and Della~Sala(2012)Niehaus, and Della~Sala]{Niehaus2012}
Niehaus,~T.; Della~Sala,~F. \emph{physica status solidi (b)} \textbf{2012},
  \emph{249}, 237\relax
\mciteBstWouldAddEndPuncttrue
\mciteSetBstMidEndSepPunct{\mcitedefaultmidpunct}
{\mcitedefaultendpunct}{\mcitedefaultseppunct}\relax
\EndOfBibitem
\bibitem[Pople et~al.(1967)Pople, Beveridge, and Dobosh]{Pople1967}
Pople,~J.~A.; Beveridge,~D.~L.; Dobosh,~P.~A. \emph{J. Chem. Phys.}
  \textbf{1967}, \emph{47}, 2026\relax
\mciteBstWouldAddEndPuncttrue
\mciteSetBstMidEndSepPunct{\mcitedefaultmidpunct}
{\mcitedefaultendpunct}{\mcitedefaultseppunct}\relax
\EndOfBibitem
\bibitem[Pople et~al.(1965)Pople, Santry, and Segal]{Pople1965}
Pople,~J.~A.; Santry,~D.~P.; Segal,~G.~A. \emph{J. Phys. Chem.} \textbf{1965},
  \emph{43}, S129\relax
\mciteBstWouldAddEndPuncttrue
\mciteSetBstMidEndSepPunct{\mcitedefaultmidpunct}
{\mcitedefaultendpunct}{\mcitedefaultseppunct}\relax
\EndOfBibitem
\bibitem[Figeys et~al.(1977)Figeys, Geerlings, and Van~Alsenoy]{Figeys1977}
Figeys,~H.; Geerlings,~P.; Van~Alsenoy,~C. \emph{Int. J. Quantum Chem.}
  \textbf{1977}, \emph{11}, 705--713\relax
\mciteBstWouldAddEndPuncttrue
\mciteSetBstMidEndSepPunct{\mcitedefaultmidpunct}
{\mcitedefaultendpunct}{\mcitedefaultseppunct}\relax
\EndOfBibitem
\bibitem[Perdew et~al.(1996)Perdew, Burke, and Ernzerhof]{perdew1996gga}
Perdew,~J.; Burke,~K.; Ernzerhof,~M. \emph{Phys. Rev. Lett.} \textbf{1996},
  \emph{77}, 3865--3868\relax
\mciteBstWouldAddEndPuncttrue
\mciteSetBstMidEndSepPunct{\mcitedefaultmidpunct}
{\mcitedefaultendpunct}{\mcitedefaultseppunct}\relax
\EndOfBibitem
\bibitem[Elstner et~al.(1998)Elstner, Porezag, Jungnickel, Elsner, Haugk,
  Frauenheim, Suhai, and Seifert]{elstner1998scc}
Elstner,~M.; Porezag,~D.; Jungnickel,~G.; Elsner,~J.; Haugk,~M.;
  Frauenheim,~T.; Suhai,~S.; Seifert,~G. \emph{Phys. Rev. B} \textbf{1998},
  \emph{58}, 7260--7268\relax
\mciteBstWouldAddEndPuncttrue
\mciteSetBstMidEndSepPunct{\mcitedefaultmidpunct}
{\mcitedefaultendpunct}{\mcitedefaultseppunct}\relax
\EndOfBibitem
\bibitem[Niehaus et~al.(2001)Niehaus, Elstner, Frauenheim, and
  Suhai]{Niehaus2001}
Niehaus,~T.~A.; Elstner,~M.; Frauenheim,~T.; Suhai,~S. \emph{J. Mol. Struct. -
  Theochem} \textbf{2001}, \emph{541}, 185--194\relax
\mciteBstWouldAddEndPuncttrue
\mciteSetBstMidEndSepPunct{\mcitedefaultmidpunct}
{\mcitedefaultendpunct}{\mcitedefaultseppunct}\relax
\EndOfBibitem
\bibitem[Aradi et~al.(2007)Aradi, Hourahine, and Frauenheim]{Aradi2007a}
Aradi,~B.; Hourahine,~B.; Frauenheim,~T. \emph{J. Phys. Chem. A} \textbf{2007},
  \emph{111}, 5678--5684\relax
\mciteBstWouldAddEndPuncttrue
\mciteSetBstMidEndSepPunct{\mcitedefaultmidpunct}
{\mcitedefaultendpunct}{\mcitedefaultseppunct}\relax
\EndOfBibitem
\bibitem[Maurice and Head-Gordon(1995)Maurice, and Head-Gordon]{Maurice1995}
Maurice,~D.; Head-Gordon,~M. \emph{Int. J. Quantum Chem. Symp.} \textbf{1995},
  \emph{29}, 361\relax
\mciteBstWouldAddEndPuncttrue
\mciteSetBstMidEndSepPunct{\mcitedefaultmidpunct}
{\mcitedefaultendpunct}{\mcitedefaultseppunct}\relax
\EndOfBibitem
\bibitem[TUR()]{TURBOMOLE}
{TURBOMOLE V6.4 2012}, a development of {University of Karlsruhe} and
  {Forschungszentrum Karlsruhe GmbH}, 1989-2007, {TURBOMOLE GmbH}, since 2007;
  available from \\ {\tt http://www.turbomole.com}.\relax
\mciteBstWouldAddEndPunctfalse
\mciteSetBstMidEndSepPunct{\mcitedefaultmidpunct}
{}{\mcitedefaultseppunct}\relax
\EndOfBibitem
\bibitem[Oddershede et~al.(1985)Oddershede, Gruener, and
  Diercksen]{Oddershede1985}
Oddershede,~J.; Gruener,~N.~E.; Diercksen,~G. H.~F. \emph{Chem. Phys.}
  \textbf{1985}, \emph{97}, 303\relax
\mciteBstWouldAddEndPuncttrue
\mciteSetBstMidEndSepPunct{\mcitedefaultmidpunct}
{\mcitedefaultendpunct}{\mcitedefaultseppunct}\relax
\EndOfBibitem
\bibitem[Krupenie(1972)]{Krupenie1972}
Krupenie,~P.~H. \emph{J. Phys. Chem. Ref. Data} \textbf{1972}, \emph{1(2)},
  423\relax
\mciteBstWouldAddEndPuncttrue
\mciteSetBstMidEndSepPunct{\mcitedefaultmidpunct}
{\mcitedefaultendpunct}{\mcitedefaultseppunct}\relax
\EndOfBibitem
\bibitem[Grabo et~al.(2000)Grabo, Petersilka, and Gross]{Gross_2000}
Grabo,~T.; Petersilka,~M.; Gross,~E. K.~U. \emph{J. Mol. Struc.-THEOCHEM}
  \textbf{2000}, \emph{501-502}, 353\relax
\mciteBstWouldAddEndPuncttrue
\mciteSetBstMidEndSepPunct{\mcitedefaultmidpunct}
{\mcitedefaultendpunct}{\mcitedefaultseppunct}\relax
\EndOfBibitem
\bibitem[Bauernschmitt and Ahlrichs(1996)Bauernschmitt, and
  Ahlrichs]{Bauernschmitt1996}
Bauernschmitt,~R.; Ahlrichs,~R. \emph{Chem. Phys. Lett.} \textbf{1996},
  \emph{256}, 454\relax
\mciteBstWouldAddEndPuncttrue
\mciteSetBstMidEndSepPunct{\mcitedefaultmidpunct}
{\mcitedefaultendpunct}{\mcitedefaultseppunct}\relax
\EndOfBibitem
\bibitem[Hesselmann and Goerling(2009)Hesselmann, and Goerling]{Hesselmann2009}
Hesselmann,~A.; Goerling,~A. \emph{Phys. Rev. Lett.} \textbf{2009}, \emph{102},
  233003\relax
\mciteBstWouldAddEndPuncttrue
\mciteSetBstMidEndSepPunct{\mcitedefaultmidpunct}
{\mcitedefaultendpunct}{\mcitedefaultseppunct}\relax
\EndOfBibitem
\bibitem[Schreiber et~al.(2008)Schreiber, Silva-Junior, Sauer, and
  Thiel]{Schreiber2008}
Schreiber,~M.; Silva-Junior,~M.; Sauer,~S.; Thiel,~W. \emph{J. Chem. Phys.}
  \textbf{2008}, \emph{128}, 134110\relax
\mciteBstWouldAddEndPuncttrue
\mciteSetBstMidEndSepPunct{\mcitedefaultmidpunct}
{\mcitedefaultendpunct}{\mcitedefaultseppunct}\relax
\EndOfBibitem
\bibitem[Jacquemin et~al.(2009)Jacquemin, Wathelet, Perpete, and
  Adamo]{Jacquemin2009}
Jacquemin,~D.; Wathelet,~V.; Perpete,~E.~A.; Adamo,~C. \emph{J. Chem. Theory
  Comput.} \textbf{2009}, \emph{5}, 2420--2435\relax
\mciteBstWouldAddEndPuncttrue
\mciteSetBstMidEndSepPunct{\mcitedefaultmidpunct}
{\mcitedefaultendpunct}{\mcitedefaultseppunct}\relax
\EndOfBibitem
\bibitem[Jacquemin et~al.(2010)Jacquemin, Perpete, Ciofini, and
  Adamo]{Jacquemin2010}
Jacquemin,~D.; Perpete,~E.~A.; Ciofini,~I.; Adamo,~C. \emph{J. Chem. Theory
  Comput.} \textbf{2010}, \emph{6}, 1532--1537\relax
\mciteBstWouldAddEndPuncttrue
\mciteSetBstMidEndSepPunct{\mcitedefaultmidpunct}
{\mcitedefaultendpunct}{\mcitedefaultseppunct}\relax
\EndOfBibitem
\bibitem[Foster and Wong(2012)Foster, and Wong]{Foster2012}
Foster,~M.~E.; Wong,~B.~M. \emph{J. Chem. Theory Comput.} \textbf{2012},
  \emph{8}, 2682--2687\relax
\mciteBstWouldAddEndPuncttrue
\mciteSetBstMidEndSepPunct{\mcitedefaultmidpunct}
{\mcitedefaultendpunct}{\mcitedefaultseppunct}\relax
\EndOfBibitem
\end{mcitethebibliography}
\end{document}